\documentclass[pra,twocolumn,showkeys,showpacs,10pt,superscriptaddress,aps]{revtex4-1}
\usepackage[english]{babel}
\usepackage{amsmath}
\usepackage{dcolumn}
\usepackage{amssymb}
\usepackage{dsfont}
\usepackage{textcomp}
\usepackage[utf8]{inputenc}
\usepackage[normal]{subfigure}
\usepackage{lipsum}
\usepackage{hyperref}
\usepackage[justification=raggedright]{caption}
\usepackage{tikz}
\usepackage{nicefrac}
\usepackage{epsfig}

\newcommand{\be}{\begin{equation}}
\newcommand{\ee}{\end{equation}}
\newcommand{\bea}{\begin{eqnarray}}
\newcommand{\eea}{\end{eqnarray}}

\renewcommand{\vec}{\mathbf}

\newcommand{\ave}[1]{\langle #1 \rangle}
\newcommand{\comut}[2]{[#1,\,#2\,]}
\usepackage{siunitx}

\usepackage{braket}
\graphicspath{{Skizzen/}}

\newcommand{\ii}{\mathrm i}

\begin{document}
	\title{Emission spectrum of broadband quantum dot superluminescent diodes}
	%from quantum dot superluminescent diodes
	
	\newcommand{\affT}{Institut für Angewandte Physik, Technische Universität Darmstadt, 64289 Darmstadt, Germany}

	\author{F. Friedrich}
	\email{Corresponding author:\\franziska.friedrich@physik.tu-darmstadt.de}
	%\affiliation{\affT}
	\author{W. Elsäßer}
	\author{R. Walser}
	\affiliation{\affT}
	\date{\today}
	
	\begin{abstract}
		We present a microscopic theory of the amplified spontaneous emission 
		of a spectrally broadband quantum dot superluminescent diode within the quantum
		white noise limit. From this multimode quantum theory, we have the ability to
		obtain all orders of temporal correlation functions. In particular, we derive
		rate equations for the optical power 
		densities, the level occupation of inhomogeneous ensemble of quantum dots 
		within the diode, as well as the emitted optical spectra. 
		As the main result, we find the external power spectrum as a convolution of the 
		intra-diode photon spectrum with a Lorentzian response.
		Assuming a Gaussian light-matter coupling results in a similar shaped Gaussian
		output  spectrum, which agrees
		very well with available experimental data.
	\end{abstract}
	
	%\pacs{42.55.Px, 42.50.Ar, 42.25.Kb}
	
	%\left\lceil 
	\maketitle
	%\tableofcontents
	
	%%%%%%%%%%%%%%%%%%%%%%%%%%%%%%%%%%%%%%%%%%%%%%%%%%%%%%%%%%%%%%%%%%%%%
	%%%%%%%%%%%%%%%%%%%%%%%%%%%%%%%%%%%%%%%%%%%%%%%%%%%%%%%%%%%%%%%%%%%%%%
	\section{Introduction}
	\label{intro}
	Commercial devices for  optical 
	coherence tomography  \cite{Velez2005,Huang1991,Judson2009} and telecommunication \cite{Urquhart2007} greatly benefit 
	from the appealing features of spectrally  broadband light-emitting quantum dot 
	superluminescent diodes (QDSLDs). In such a QDSLD, light is generated in the 
	regime of amplified spontaneous emission (ASE), i.\thinspace{}e. right at the
	border 
	between spontaneous and stimulated emission.
	Careful design of waveguide geometry and gain medium results in very high output
	intensities, spatial coherence and a very broad radiation spectrum width in the
	THz regime 
	\cite{Siegman1986,Wiersma2008}, 
	thus combining spatial coherence features of typical laser 
	diodes with the remarkable broadband spectrum of light-emitting diodes.

	Modern applications like ghost imaging  
	\cite{pittman1995,Kuhn2016,Hartmann2015,Seppel2017} or the analysis of
	fundamental 
	quantum optical questions on the photon statistics 
	\cite{Boitier2010,Boitier2011,Blazek2011b,Blazek2012,Boitier2013} 
	benefit from  these semiconductor devices. In 2011, Blazek {\em et al.} \cite
	{Blazek2011a} measured temporal second-order correlation function on the fs time
	scale with fast 
	two-photon detectors \cite{Boitier2009}. 
	They discovered a novel state of light, which features a reduction of the
	second-order 
	degree  of coherence $g^{(2)}(\tau=0)$ from 2 to 1.33 at a temperature of 
	\SI{190}{\kelvin} without any spectral narrowing (spectral width remains 
	\si{\tera\hertz}). Thus, these novel states of light are simultaneously
	incoherent in 
	first- and coherent in second-order correlation function. Understanding this
	'hybrid coherent light' is an interesting topic and provides more insights into
	the quantum nature of light-emitting broadband semiconductor sources.  
	
	Originally, amplified spontaneous emission has been already studied by 
	Allen and Peters \cite{Peters1971a,Peters1971b,Peters1971c,Peters1972d} in 
	1971. Amongst others, they formulated the ASE threshold condition and 
	investigated spatial coherence properties of a classical field propagating 
	in a bulk medium in the absence of an external cavity. 
	The global spreading of LED illumination technology has revived the 
	interest in the radiation properties of diodes, for example studies of the 
	emission and photon statistical characteristics of QDSLDs 
	\cite{Jechow2013,Jechow2017}. 
	Numerical models have been developed based on rate equations 
	\cite{gioa,BardellaRossettiMontrosset2009}, 
	traveling wave approaches 
	\cite{RossettiBardellaMontrosset2011} 
	and finite element methods \cite{Li2010}. 
	
	Microscopic theories describing ASE are mostly focused on the specific 
	semiconductor properties \cite{Poel2005,Wegert2011}. However, hybrid 
	coherent light generation is an optical process and should not 
	require detailed information about semiconductor charge carrier dynamics.
	Thus, a quantum optical analysis of ASE adapted to 
	superluminescent diodes with tilted end facets using a quantum dot gain 
	medium is appropriate.
	
	In a previous study of the optical spectral properties of such QDSLDs, a 
	multimode phase-randomized Gaussian state was postulated as a premise to 
	match the observed spectral shape \cite{HartmannFriedrich2015}. 
	One finds that the second-order degree of coherence is inversely proportional to
	the number of available modes, i.\thinspace{}e. $g^{(2)}(\tau)$ is affected by
	the characteristics of each individual QDSLD. This prediction was experimentally
	confirmed by an optical feedback experiment
	\cite{Hartmann2013blazek,HartmannFriedrich2015}. 
	Further evidence on the nature of the optical quantum state came from an 
	experiment superimposing a coherent single-mode laser beam with the broadband
	radiation 
	of the QDSLD. The observed spectra agree with the theoretical analysis very
	well.
	
	In order to shed more light on the nature of the photonic state, we will discuss
	in this article a microscopic model of the QDSLD, based on the input-output
	formalism of laser theory \cite{sargent1976,quantumnoise}. Based on this
	multimode theory, we determine temporal first-order 
	correlation function of the radiation field measured by a single-photon detector
	and compare them with the experimental data.
	
	The article is organized as follows: after introducing basic facts of QDSLDs in
	Sec.~\ref{ExpFacts}, we establish a theoretical model, where we consider the
	energy of the intrawaveguide system and set up the Ito quantum stochastic
	differential equations for the diode system. Within the rate equation limit, we
	study  in Sec.~\ref{model} the stationary photon number as a function of the
	incoherent pumping rate for the special cases: 1) a single ASE  field mode and
	2) a multimode radiation  field, both coupled to a homogeneous ensemble  of
	quantum dots. In Sec.~\ref{output}, we use the input-output formalism to analyze
	the first-order autocorrelation function directly related to the output spectral
	density of light-emitting QDSLDs. A conclusion is given in
	Sec.~\ref{conclusion}. Three appendices contain details on the quantum Ito
	calculus in App.~\ref{itoQSDE},
	the multi-channel transmission spectrum of a passive waveguide in
	App.~\ref{emptyCavity}
	as well as the interpolation procedure of the power spectrum in
	App.~\ref{Fitparam}.
	%%%%%%%%%%%%%%%%%%%%%%%%%%%%%%%%%%%%%%%%%%%%%%%%%%%%%%%%%%%%%%%
	\section{Properties of quantum dot superluminescent diodes}
	\label{ExpFacts}
\begin{figure}[h!]
		\includegraphics[width=1\columnwidth]{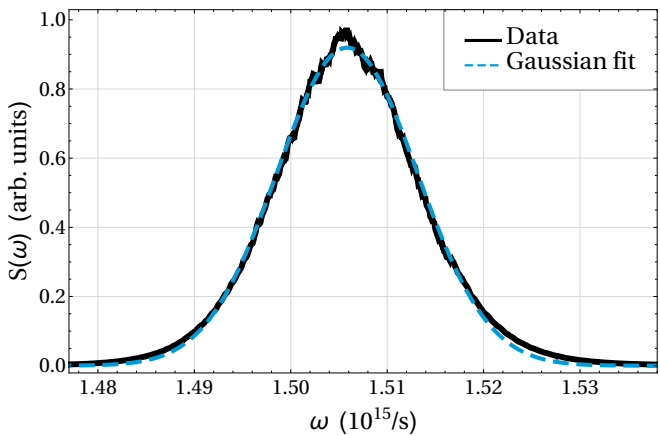}
		\caption{Typical shape of an experimental QDSLD power spectrum 
			$S(\omega)$ (black, solid line) versus angular frequency $\omega$. 
			Central frequency 
			$\bar\omega=2\pi \cdot\SI{0.24}{\peta\hertz}$ ($\bar
			\lambda=\SI{1249.14}{nm}$), standard deviation $\sigma=2\pi\cdot
			\SI{1.16}{\tera\hertz}$  and spectral width $b=2\pi\cdot \SI{4.11}{THz}$
			($\Delta \lambda=\SI{21.39}{nm}$) are obtained from a Gaussian fit 
			(blue, dashed line, Eq.~(\ref{gaussfit})).}
		\label{spectrum}
	\end{figure}

	Superluminescent diodes \cite{LeeBurrusMiller1973,Amann1979} are optoelectronic
	semiconductor devices emitting spatially directed light with spectral widths in
	the \si{\tera\hertz} regime. 
	A typical measured optical power spectrum \footnote{We thank S\'{e}bastien
		Blumenstein (name of birth Hartmann) for providing the experimental data.}  of a
	QDSLD 
	is shown in Fig.~\ref{spectrum}, where experimental
	data are compared with a Gaussian fit (Eq.~(\ref{gaussfit})).
	
	In the setup of Ref.~\cite{Blazek2011a}, a tilted optical 
	waveguide was used (angle 5-8° with respect to the longitudinal axes). 
	The emission facets where anti-reflection coated to 
	suppress the formation of longitudinal modes~\cite{Alphonse1988}. 
	The gain material of a QDSLD consists of quantum dots differing  in size, shape
	and material composition, leading to a considerable inhomogeneous
	level-broadening. In order to obtain high efficient light amplifications with
	broad 
	spectral ranges \numrange{5}{10} layers of the 
	gain medium are essential, where each layer consists of about 
	400 quantum dots\si{\per \micro \meter^{2}}. They are spatially separated by
	other 
	semiconductor materials, the so-called buffer layers having a refractive 
	index, $n_m=3.5$, slightly smaller than the index of the single gain layers,
	$n_c =3.505$ \cite{BlazekPrivatCom}. 
	%%%%%%%%%%%%%%%%%%%%%%%%%%%%%%%%%%%%%%%%%%%%%%%%%%%%%%%%%%%%%%%
	\section{Model of a quantum dot superluminescent diode}
	\label{model}
\begin{figure}[h!]
		\includegraphics[width=1\columnwidth]{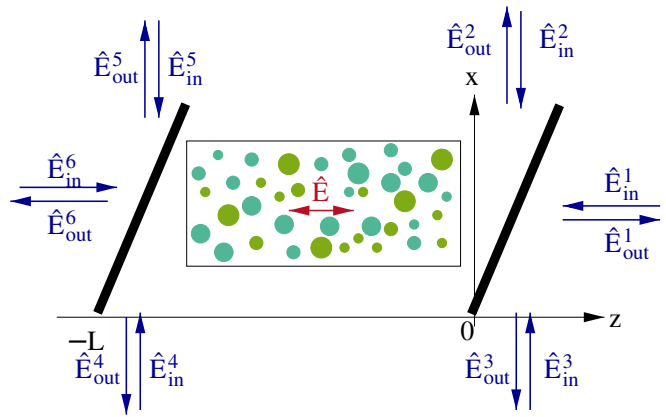}
		\caption{Model of a QDSLD (top view): bulk material in a 
			rectangular waveguide hosts $M$ quantum dots. 
			The tilted end facets of the QDSLD prevent reflection 
			back into the medium and are formally represented by beam splitters.
			In principle, six external quantum channels  
			$\hat{\vec E}^{\alpha}$ with $\alpha=1,...,6$ couple into the waveguide and
			interact with the ASE field 
			$\hat {\vec E}$.}
		\label{sld}
	\end{figure}

	%%%%%%%%%%%%%%%%%%%%%%%%%%%%%%%%%%%%%%%%%%%%%%%%%%%%%%%%%%%%%%%%%%%%%%%%%%%%%%%%
	Let us consider a simple diode model with one layer of gain 
	medium, as depicted in Fig.~\ref{sld}. There, the gain medium consists of 
	$M$ quantum dots, which are embedded in a  bulk material that defines a
	waveguide. 
	%\begin{figure}[h!]
	%  \centering
	% \scalebox{.59}{\input{pictures/sld_xfig/SLDtest.pdf_t}}
	%    \caption{Model of a QDSLD: absorptive bulk material in a 
	%		rectangular waveguide hosts $M$ quantum dots. 
	%		The tilted end facets of the QDSLD prevent reflection 
	%back into the medium and are formally represented by beam splitters.
	%		In principle, six external quantum channels  
	%$\hat{\vec E}^{\alpha}$ with $\alpha=1,...,6$ couple into the waveguide and
	%interact with the ASE field 
	%		$\hat {\vec E}$.}
	%  \label{sld}
	%\end{figure}
	%%%%%%%%%%%%%%%%%%%%%%%%%%%%%%%%%%%%%%%%%%%%%%%%%%%
	\subsection{ASE field}
	We now study the broadband light  inside the diode which must be 
	described by  a multimode electrical field, 
	$\hat{\vec{E}}=\hat{\vec{E}}^{(+)}+\hat{\vec{E}}^{(-)}$
	with positive frequency part 
	\begin{align}
	\label{ase}
	\hat{\vec{E}}^{(+)}(\vec r,t)&=
	% \hat{\vec{E}}^{(+)}(x,y,t-\frac{z}{c})=
	\sum_{\{k_i\}}  \vec u_i(\vec r)\hat a_i(t).
	\end{align}
	As we want to pursue a quantum theory of light, we need to consider a Hamilton
	operator
	\begin{align}
	\label{fieldhamiltonianase}
	\hat{\mathcal{H}}_r=\sum_{\{k_i\}}\hbar\omega_{i}\hat a_i^\dagger\hat a_i^{ },
	\end{align}
	depending on the bosonic amplitudes $\hat a_i$, which satisfy the commutation
	relation
	$\comut{\hat a_i^{ ^{ }}}{\hat a_{j}^\dagger}=\delta_{ij}$.
	The electric field in Eq.~(\ref{ase})   is formed by a superposition of $N$
	modes of type
	$\vec u_i(\vec r)= \mathcal{E}_i\, \chi(x,y)e^{\ii k_i z} \vec e_y$,
	which are solutions of the Helmholtz equation
	\begin{align}
	\label{helmholtz}
	\left(\Delta+k_i^2\right)\vec u_i(\vec r)=0,
	\quad
	\omega_i=|k_i| \frac{c}{n_c}
	\end{align}
	with rectangular geometry.
	They factorize into a single transverse wave function $\chi(x,y)$ and 
	longitudinal plane waves with wave numbers $|k_i|$ with 
	$k_i=2\pi i n_c/L$. The length of the waveguide is denoted by $L$.
	Due to the rectangular geometry of the QDSLD \cite{zhang2007effect}, the 
	field is linearly polarized in $y$-direction. The transverse 
	mode function is normalized to the cross-section area 
	$A=\int_A d^2x \,|\chi|^2$. If the system volume is $V=AL$, 
	then the normalization factor of the electric field reads  
	$\mathcal{E}_i=\ii\sqrt{\hbar \omega_i/2\epsilon V}$. 
	According to Eq.~\eqref{helmholtz}, there is a linear dispersion relation 
	between frequency and wave number with vacuum speed of light $c$.
	%%%%%%%%%%%%%%%%%%%%%%%%%%%%%%%%%%%%%%%%%%%%%%%%%%%%%%%%%%%%%%
	\subsection{External fields}
	%Coupling into or out-of a tilted end facet with anti-reflection coating 
	%\cite{Blazek2011a} is formally equivalent to the 
	%four-port beam splitter, sketched in 
	%Fig.~\ref{bs}. Output amplitudes are connected to input amplitudes
	%by the S-matrix \cite{vogel2006quantum},
	%\begin{gather}
	%	\hat{\vec  b}_\text{out}(\omega)=S(\omega)~\hat{\vec b}_\text{in}(\omega),
	%\end{gather}
	%which depends on the reflectance $\mathcal R$ and transmittance~$\mathcal T$ of
	%the beam splitter device.
	%%
	%\begin{figure}[h!]
	%  \centering
	%    \scalebox{.45}{\input{pictures/BS/bs.pdf_t}}
	%    \caption{Sketch of a fourport beam splitter.}
	%  \label{bs}
	%\end{figure}
	The input-output arrangement of the diode system depicted in Fig.~\ref{sld}
	couples the internal ASE field 
	$\hat{\vec E}$ to the outside. 
	There, six outside channels $\alpha \in \{1,...,6\}$
	\begin{align}
	\hat{\vec E}^{\alpha(+)}(\vec r,t)=\sum_{\{k_i \}}\sum_{q\in M_i^\alpha}\vec
	v^\alpha_q(\vec r) \hat b_{iq}^\alpha(t),
	\end{align} 
	are themselves multimode fields
	with amplitudes $\hat b^{\alpha}_{iq}$ oscillating at frequency
	$\omega_q=c|k_q|$ of frequency group $M_i^\alpha$. The corresponding Hamilton
	operator is given by
	\begin{align}
	\label{hambath}
	\hat{\mathcal H}_{c}=\sum_{\alpha=1}^6 \sum_{\{k_i\}}\sum_{q\in M_i^\alpha}
	\hbar \omega_q \hat b_{iq}^{\alpha\dagger} \hat b_{iq}^\alpha.
	\end{align}
	%For simplicity, we assume that the electromagnetic field outside of the diode
	%is captured by single-mode fibers, which support
	%linearly polarized, single modes  
	For simplicity, we neglect  spatial expansion of the electromagnetic field
	emitted by the diode  and consider only a one-dimensional 
	propagation. Clearly, from an experimental point of view this can be achieved by
	single-mode fibers, which support
	linearly polarized, single modes
	\begin{align}
	\vec v^\alpha_q(\vec r)&=\mathcal E_q\vec e_y 
	\begin{cases}
	\xi(x,y)e^{\mathrm{i} k_q z} & \alpha=1,6\\
	\xi(y,z)e^{\mathrm{i} k_q x} & \alpha=2,3,4,5\\
	\end{cases},
	\\
	\label{normcoef}
	\mathcal E_q&=\ii \sqrt{\frac{\hbar \omega_q}{2\epsilon_0 V}} 
	\end{align}
	as well as a transverse mode function $\xi$ that is normalized to the
	cross-section area $A=\int_A d^2x \,|\xi|^2$.
	We define frequency groups of outside modes 
	\begin{align}
	M_{i}^{\alpha}=\{q\in\mathbb{Z}:q^\text{min}_\alpha(k_i)\le q\le q^
	\text{max}_\alpha(k_i)\},
	\end{align}
	with equidistant frequency spacing $\Delta\omega=2\pi \Delta\nu$
	which couple 
	to individual ASE field modes at 
	frequency $\omega_i$ in the sense of well resolved bands. 
	To be specific, we specify the set of modes  
	which couple an ASE mode $i$ to 
	$\hat{b}_{iq}^{\alpha}(t)$, $q\in M_{i}^{\alpha}$,
	which satisfy the bosonic commutation relation 
	\begin{align}
	\comut{\hat b_{iq}^{\alpha}(t)}{\hat b_{jp}^{\beta\dagger}(t)}=\delta_{\alpha 
		\beta}\delta_{qp}\delta_{ij}.
	\end{align}
	Here, correlations between the channel fields are excluded within our 
	theoretical considerations.
	
	Assuming that single frequencies $\omega_q$ are centered around a certain
	frequency $\omega_i$, that is $\omega_q\approx \omega_i$,  so that $\mathcal
	E_q\approx \mathcal E_i$ (see Eq.~\eqref{normcoef}) and 
	introducing the free input- and output fields,
	\begin{align}
	\label{bi}
	\hat b_{i,\text{in}}^{\alpha}(t)&
	\equiv\sqrt{\Delta\nu}
	\sum_{q\in M_{i}^\alpha}
	e^{-\mathrm i \omega_q(t-t_i)}
	\hat b_{iq}^{\alpha}(t_i), \\
	\label{bo}
	\hat b_{i,\text{out}}^{\alpha}(t)&
	\equiv\sqrt{\Delta \nu}
	\sum_{q\in M_{i}^\alpha}
	e^{-\mathrm i\omega_q(t-t_f)}
	\hat b_{iq}^{\alpha} (t_f),
	\end{align}
	in terms of their initial and final values at $t_i$ and $t_f$, 
	respectively, the positive frequency part of the electric field from channel 1
	emitted on the right end facet  ($z>0$) reads \cite{quantumnoise}
	\begin{gather}
	\label{channelfield}
	\hat{\vec{E}}^{1(+)}(\vec{r},t)
	=\hat{\vec{E}}_{\text{in}}^{1(+)}(x,y,\tau^{-})
	+\hat{\vec{E}}_{\text{out}}^{1(+)}(x,y,\tau^+),
	\\
	\label{einout}
	\hat{\vec{E}}_{\text{in/out}}^{1(+)}(x,y,\tau)=
	\sum_{\{k_i>0\}} \vec v_i^1(x,y) 
	\hat{b}_{i,\text{in/out}}^{1}(\tau),
	\end{gather}
	with $	\tau^{\pm} = t \pm\nicefrac{z}{c} $ and $\vec v_i^1(x,y)= \mathcal
	E_i\xi(x,y)\vec e_y/\sqrt{\Delta\nu}$.
	The other optical channels $\hat{\vec E}^\alpha$ of Fig.~\ref{sld} can be
	described in an analogous fashion, but are not required for this discussion.  
	
	Evaluating the frequency sum in the continuum limit 
	(c.f. App. \ref{emptyCavity}), requires the frequency spacing 
	$\Delta\nu$ as an integration measure. 
	This procedure defines the "quantum white noise" limit \cite{Gardiner1985},
	\begin{equation}
	\comut{\hat b_{i ,\text{in/out}}^{\alpha}(t)}{
		\hat b_{j,\text{in/out}}^{\beta\dagger}(t')}=
	\delta_{ij}\delta_{\alpha \beta}\delta(t-t'),
	\label{eq:}
	\end{equation}
	for the input and output fields and demonstrates their infinitely 
	short correlation time.
	%%%%%%%%%%%%%%%%%%%%%%%%%%%%%%%%%%%%%%%%%%%%%%%%%%%%%%%%%%%%%%%
	\subsection{Gain medium}
\begin{figure}[h]
		\centering
		\includegraphics[width=.6\columnwidth]{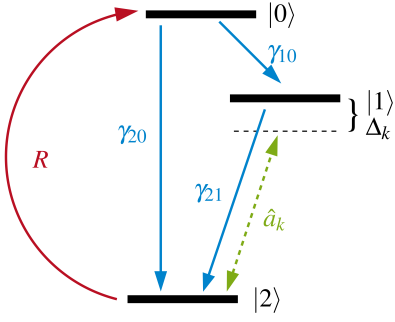}
		\caption{Three-level quantum dot energy diagram with decay rates
			$\gamma_{ij}$ $(i,j\in\{0,1,2\})$, driven by the ASE radiation field with
			amplitudes $\hat a_i$ and incoherently pumped at rate $R$.}
		\label{qd}
	\end{figure}

	The gain medium is formed by an ensemble of $M$ quantum dots (QDs) embedded in 
	semiconductor material. In order to achieve light amplification, we need to 
	consider at least three levels \cite{MIMCSB2000}.
	The Hamiltonian of the QD system reads
	\begin{align}
	\label{hamiltonianqd}
	\hat{\mathcal H}_{a}=\sum_{j=1}^M \sum_{r=0}^2\hbar \omega_r^j
	\hat{\sigma}_{rr}^{j\dagger} \hat{\sigma}_{rr}^j.
	\end{align}
	$\hat \sigma_{rr}^j$ denotes the population of level $\ket{r}$ of the $j$th
	quantum dot.
	In Fig.~\ref{qd}, we show a closed three-level system 
	with eigen-energies $\hbar \omega_i$ and 
	transition frequencies $\omega_{ij}=\omega_i-\omega_j$ between 
	level $i$ and $j$. 
	% \begin{figure}[h]
	%  \centering
	%    \scalebox{.45}{\input{pictures/4levelatom_xfig/4levelatom.pdf_t}}
	%    \caption{Three-level quantum dot energy diagram with decay rates
	%$\gamma_{ij}$ $(i,j\in\{0,1,2\})$, driven by the ASE radiation field with
	%amplitudes $\hat a_i$ and incoherently pumped with rate $R$.}
	%  \label{qd}
	%\end{figure}
	
	Each quantum dot is incoherently pumped with rate $R$ and releases the energy 
	via amplified spontaneous emission on the $\ket{1}\to \ket{2}$ transition to the
	multimode ASE field \eqref{ase}. 
	%The ASE field propagating in the semiconductor bulk material experiences a high
	%rate of absorption in direct 
	%competition with the gain.
%	The ASE field propagating in the semiconductor bulk material experiences losses
%	in direct 
%	competition with the gain. This can be modeled by coupling quantum dot $j$ to a
%	single bath of energy
	We couple each quantum dot $j$ to a single phonon bath of energy
	\begin{align}
	\hat{\mathcal H}_\text{b}^j=
	\sum_{\{s\}} \sum_{\{k\}}\hbar\omega_{s k}^j\hat{b}^{j\dagger}_{
		s k}\hat{b}_{s k}^j,
	\quad s\in \{(21),(20),(10)\},
	\end{align}
	given by the sum of three independent baths, each acting on one particular 
	quantum dot transition $s$. They satisfy the bosonic commutation relation 
	\begin{align}
	\comut{\hat b^{j}_{s k}(t)}{\hat b^{j\dagger}_{s' k'}(t)}=\delta_{ k
		k'}\delta_{ss'}.
	\end{align}
%	Regarding the absorption coefficient of solids, interactions between quantum
%	dots are assumed to be not possible.
Finally, we disregard  interactions between quantum dots.
	%%%%%%%%%%%%%%%%%%%%%%%%%%%%%%%%%%%%%%%%%%%%%%%%%%%%%%%%%%%%%%%
	\subsection{Light-amplification by inhomogeneous broadened quantum dots}
	\label{energy}
	The energy of the isolated diode system, 
	\begin{align}
	\hat{\mathcal{H}}_s=\hat{\mathcal{H}}_r+\hat{\mathcal{H}}_a+\hat{\mathcal{H}}_{\text{int}},
	\end{align}
	is given by the sum of the energy of the ASE field modes
	\eqref{fieldhamiltonianase},
	the energy of the $M$ atomic system \eqref{hamiltonianqd}
	and the dipole interaction between the broadband light field and the $M$
	distinguishable quantum dots in the rotating wave approximation,
	\begin{align}
	\hat{\mathcal{H}}_{\text{int}}=-\ii\sum_{\{k_i\}} \sum_{j=1}^M \hbar
	g_i^j\hat{\sigma}_ {21}^{j\dagger}\hat{a}_i^{ }+\text{h.c.},
	\end{align}
	with transition operator $\hat \sigma_{21}^j$ between level $\ket{2}$ and
	$\ket{1}$.
	The coupling constant,
	$g_i^j=-\ii \vec{d}_{21}^{j\ast}\vec u_i(\vec{r}_j)/\hbar$, depends  on
	frequency $\omega_i$ and is proportional to the dipole matrix element
	$\vec{d}_{21}^j$.
	
	Switching to a suitable interaction picture 
	\begin{align}
	\label{Heisenberg}
	\hat{H}_s=\ii\hbar \dot{\hat U}^\dagger\hat U+\hat U^\dagger\hat {\mathcal H}_s\hat U,
	\end{align}
	with transformation operator 
	\begin{align}
	\hat{U}=\text{exp}(-\ii\hat Gt),
	\end{align}
	where
	\begin{align}
	\begin{split}
	% \hat{U}=e^{-\ii\hat Xt} \quad \text{with} \quad
	\hat G=&\sum_{\{k_i\}}\bar \omega_{12}\hat{a}_i^\dagger\hat{a}_i^{ }
	+\sum_{j=1}^M
	[\omega_0^j\hat{\sigma}_{00}^j-(\omega_{1}^j-\delta\omega_{12}^j)\hat{\sigma}_{11}^j
	+\omega_2^j\hat{\sigma}_{22}^j ],
	\end{split}
	\end{align}
	the time dependent Hamiltonian reads
	\begin{equation}
	\label{systemhamiltonian}
	\begin{split}
	\hat{H}_s=&\hbar \sum_{\{k_i\}}  \Delta_i \hat{a}_i^\dagger \hat{a}_i^{
	}
	+\hbar\sum_{j=1}^M  \delta \omega_{12}^j \hat{\sigma}_{11}^j 
	\\&
	-\mathrm i \hbar \sum_{\{k_i\}} \sum_{j=1}^M (
	g_i^j\hat{\sigma}_{21}^{j\dagger}\hat{a}_i + \text{h.c.} ).
	\end{split}
	\end{equation}
	In this interaction  picture, the ASE field oscillates with detuning
	$\Delta_i=\omega_i-\bar \omega_{12}$, where $\bar \omega_{12}=\sum_{\{k_i\}}
	\omega_i/N$ is the mean transition frequency of the quantum dots. Accordingly,
	the frequency $\delta \omega_{12}^j=\omega_{12}^j-\bar \omega_{12}$ represents a
	small deviation of the $j$th quantum dot from the mean value.   
	%%%%%%%%%%%%%%%%%%%%%%%%%%%%%%%%%%%%%%%%%%%%%%%%%%%%%%%%%%%%%
	\subsection{Intrawaveguide field}
	\label{intracavity}
	%%%%%%%%%%%%%%%%%%%%%%%%%%%%%%%%%%%%%%%%%%%%%%%%%%%%%%%%%%%%%   
	In order to describe the total QDSLD intrawaveguide system, we follow the
	concepts of C. Gardiner in 1985 \cite{Gardiner1984,Gardiner1985} dealing with
	open quantum systems. Using the Ito form of a quantum stochastic differential
	equation (QSDE) defined in \eqref{qsde}, the differential of the $i$th ASE field
	mode in terms of the input noise operators reads
	\begin{align}
	\label{QSDEase}
	\begin{split}
	d\hat{a}_i\!=&\!-\ii[\Delta_i -\ii\gamma_i^{lr}] \hat a_i dt +\sum_{j=1}^M
	g_i^{j\ast}\hat{\sigma}_{21}^jdt
	\\&
	-d\hat{B}_{i}^{l}-d\hat{B}_{i}^{r},
	\quad
	\gamma_i^{lr}=\frac{1}{2}(\gamma_i^l+\gamma_i^r).
	\end{split}
	\end{align}
	The first term in Eq.~\eqref{QSDEase} describes the free evolution of the mode
	$i$ and includes the external damping rates $\gamma_i^l$, $\gamma_i^r$ resulting
	from the left and right input noise fields leading to damping effects of the
	system. The second term describes the interaction of the ASE field mode with all
	$M$ quantum dots and the last two terms consider noise fluctuations.
	According to Fig.~\ref{sld}, the left and right noise operators are given by 
	\begin{align}
	&d\hat B_{i}^{r}(t)
	=\frac{\zeta_i^{r\ast}}{2}(d\hat{b}_{i}^{1}+d\hat{b}_{i}^{2}),
	\,\,\, d\hat B_{i}^{l}(t)
	=\frac{\zeta_i^{l\ast}}{2}(d\hat{b}_{i}^{4}+d\hat{b}_{i}^{6}),
	\end{align}
	with 
	\begin{align}
	\label{zeta}
	\zeta_i^{\alpha}=\sqrt{\gamma_i^\alpha}~ e^{\ii \phi_i^\alpha}
	\end{align}
	and normalized Ito increments $db_i^\alpha$, ($\alpha=1,...,6$) satisfying the
	commutation relation,
	\begin{align}
	\begin{split}
	\comut{d\hat b_{i}^{\alpha}(t)}{d\hat b_{j}^{\beta\dagger}(t')}
	=\delta_{ij} \delta_{\alpha\beta} \delta(t-t') dt dt',
	\end{split}
	\end{align}
	%Obviously, the Ito increments have the dimension of the square root of 
	%reciprocal time and satisfy
	as well as the Ito algebra listed in App. \ref{itoQSDE}. The damping rates on
	the left and right side of the beam splitter are defined by 
	\begin{align}
	\label{dampingrate}
	\gamma_i^r\equiv 2\gamma_i^1=2\gamma_i^2, 
	\quad
	\gamma_i^l\equiv2\gamma_i^4=2\gamma_i^6.
	\end{align}  
	Please note, the input fields of channel 3 and 5,
	$\hat{\vec{E}}^3_{\text{in}}$ and $\hat{\vec{E}}^5_{\text{in}}$, never enter the
	waveguide. Therefore, the corresponding damping rates, proportional to the
	coupling strength between reservoir and ASE field, are equal to zero.
	Furthermore, we claim that all phases $\phi_i^\alpha$ of the channels from the
	left and right hand side are equal, i.e. 
	\begin{align}
	\label{phase}
	\phi_i^r\equiv\phi_i^1=\phi_i^2, \quad  \phi_i^l\equiv\phi_i^4=\phi_i^6.
	\end{align}
	In the same way, we can derive the QSDE for the atomic coherences and
	populations of the quantum dots.
	Introducing a global incoherent pumping rate  $R=n_{20}^j\gamma_{20}^j$
	\cite{Tian1992,Tawara2010,Sturm2014} acting equally strong on each individual
	quantum dot  (see Fig.~\ref{qd}), the quantum dot populations and coherences in
	terms of the Ito increments $d\hat b_m^j=d\hat B_m^j/\sqrt{\gamma_m^j}$ read:
	\\
	\\
	Populations:
	\begin{align}
	\begin{split}
	d\hat \sigma_{00}^j=&
	\bigl[-\Gamma_{0}^j\hat \sigma_{00}^j
	+\gamma_{10}^j n_{10}^j \hat \sigma_{11}^j
	+R \hat \sigma_{22}^j \bigr] dt 
	\\&
	-\Bigl(\hat \sigma_{10}^{j\dagger}d\hat B_{10}^j+\hat
	\sigma_{20}^{j\dagger}d\hat B_{20}^j +\text{h.c.}\Bigr) 
	\end{split}
	\end{align}
	\begin{align}
	\begin{split}
	d\hat \sigma_{11}^j=&
	-\Gamma_{1}^j\hat \sigma_{11}^j dt
	-\sum_{\{k_i\}}( g_i^{j\ast}\hat a_i^\dagger\hat \sigma_{21}^j dt
	+\text{h.c.} ) \\
	&
	+\gamma_{10}^j(n_{10}^j+1)\hat \sigma_{00}^j dt
	+\gamma_{21}^jn_{21}^j \hat \sigma_{22}^j dt  \\
	&-(\hat \sigma_{21}^{j\dagger}d\hat B_{21}^j-\hat \sigma_{10}^{j\dagger}d\hat
	B_{10}^j
	+\text{h.c.})
	\end{split}
	\end{align}
	\begin{align}
	\begin{split}
	d\hat \sigma_{22}^j=&
	-\Gamma_{2}^j\hat \sigma_{22}^j dt
	+\sum_{\{k_i\}} (g_i^{j\ast}\hat a_i^\dagger\hat \sigma_{21}^j dt
	+\text{h.c.})  
	\\&
	+\gamma_{21}^j(n_{21}^j+1)\hat \sigma_{11}^j dt
	+\bigl(R+\gamma_{20}^j\bigr)\hat \sigma_{00}^j dt  \\
	&+\left(\hat \sigma_{21}^{j\dagger}d\hat B_{21}^j+\hat
	\sigma_{20}^{j\dagger}d\hat B_{20}^j+\text{h.c.}\right) 
	\end{split}
	\end{align}
	\\
	and coherences:
	\begin{align}
	\label{coherence}
	\begin{split}
	d\hat \sigma_{21}^j=&
	-\bigl[\ii(\delta\omega_{12}^j-\ii\Gamma_{21}^j)\hat\sigma_{21}^j+
	\sum_{\{k_i\}} g_i^j\hat w^j\hat a_i
	\bigr]dt
	\\
	&
	+\hat w^jd\hat B_{21}^j  
	+d\hat B_{10}^{j\dagger} \hat \sigma_{20}^j 
	+\hat \sigma_{10}^{j\dagger}~d\hat B_{20}^j
	\end{split}
	\end{align}
	\begin{align}
	\begin{split}
	d\hat \sigma_{20}^j=&
	\bigl[-\Gamma_{20}^j \hat \sigma_{20}^j 
	+\sum_{\{k_i\}} g_i^j\hat \sigma_{10}^j \hat a_i \bigr] dt
	\\
	&+ \hat \sigma_{10}^j~d\hat B_{21}^j 
	-\bigl(\hat \sigma_{22}^j-\hat \sigma_{00}^j\bigr) d\hat B_{20}^j
	-\hat \sigma_{21}^j  d\hat B_{10}^j  
	\end{split}
	\end{align}
	\begin{align}
	\begin{split}
	d\hat \sigma_{10}^j =&
	\bigl[\ii(\delta\omega_{12}^j-\ii\Gamma_{10}^j)\hat \sigma_{10}^j
	-\sum_{\{k_i\}} g_i^{j\ast}\hat a_i^\dagger \hat \sigma_{20}^j \bigr]dt
	\\
	&-d\hat B_{21}^{j\dagger}\hat \sigma_{20}^j
	-\hat \sigma_{21}^{j\dagger} d\hat B_{20}^j 
	-\bigl(\hat \sigma_{11}^j-\hat \sigma_{00}^j\bigr) d\hat B_{10}^j
	\end{split}
	\end{align}
	with inversion $\hat w^j=\hat\sigma_{11}^j-\hat\sigma_{22}^j$ and  decay rates
	\begin{align}
	\Gamma_{0}^j&=\gamma_{10}^j(n_{10}^j+1)+\gamma_{20}^j+R,\\
	\Gamma_{1}^j&=\gamma_{21}^j(n_{21}^j+1)+\gamma_{10}^jn_{10}^j,\\
	\Gamma_{2}^j&=\gamma_{21}^jn_{21}^j+R,\\
	\Gamma_{mn}^j&=\frac{1}{2}\bigl(\Gamma_{mm}^j+\Gamma_{nn}^j\bigr)
	,
	\,\,\,
	m,n \in \{0,1,2\},
	\,\,\,
	m \neq n.
	%   \Gamma_{21}^j&=\frac{1}{2}\bigl(\Gamma_{11}^j+\Gamma_{22}^j \bigr),\\
	%   \Gamma_{20}^j&=\frac{1}{2}\bigl(\Gamma_{00}^j+\Gamma_{22}^j),\\
	%   \Gamma_{10}^j&=\frac{1}{2}\bigl(\Gamma_{00}^j+\Gamma_{11}^j) .
	\end{align}
	%%%%%%%%%%%%%%%%%%%%%%%%%%%%%%%%%%%%%%
	\subsection{On the nature of the quantum state of the QDSLD system}
	\label{assumptions}
	Regarding the spectral measurements of highly incoherent broadband radiation
	fields (see Fig.~\ref{spectrum}), we assume a missing coherent amplitude of ASE
	mode $i$, that is
	\begin{align}
	\ave{\hat a_i}&=\text{Tr}\{\hat a_i\hat \rho\}=0.
	\end{align}
	Furthermore, we disregard correlations between different ASE field
	modes at the same space-time event, so that the average value,
	\begin{align}
	\ave{\hat a_i^\dagger \hat a_{j}^{ }}&=n_i \delta_{ij},
	\end{align}
	is given by the photon number $n_i$ of mode $i$.
	Regarding the input noise field, we do not allow interactions between the
	different external channel modes and between these 
	modes with the  quantum dot bath modes inside the waveguide.
	Occupation numbers and decay rates of the left and right hand
	side of the waveguide are assumed to be equal for all ASE field modes, that is
	$n_i^\alpha= n^\alpha$, $\gamma_i^\alpha= \gamma^\alpha$ with $\alpha=l,r$.
	
	As the influence of a single quantum dot on the ASE radiation field mode $i$ is small, higher order corrections can be neglected by
	decorrelating field and atomic operator (decorrelation approximation)
	\cite{barnett2002}, i.e.
	\begin{align}
	\ave{\hat a_{m}^\dagger\hat w^j\hat a_n^{ }}
	\approx \ave{\hat a_{m}^\dagger\hat a_{n}^{ }} \ave{\hat w^j}.
	\end{align}
	Furthermore, we adiabatically eliminate coherences as dynamical variables in
	the averaged QSDEs by replacing $\ave{\hat{\sigma}_{21}^{j\dagger} \hat a_i}$ by
	its stationary solution
	\begin{align}
	\label{decorr}
	\ave{\hat \sigma_{21}^{j\dagger} \hat a_i}
	\approx
	\frac{g_i^{j\ast}\bigl(n_i \ave{\hat
			w^j}+\ave{\hat{\sigma}_{11}^j}\bigr)}{\ii\Delta_i^{j}+\Gamma^j},
	\end{align}
	with photon number $n_i$ of mode $i$, $\Gamma^j\equiv\Gamma_{21}^j+\gamma^{lr}$
	with $\Gamma_{21}^j$ defined in Eq.~\eqref{coherence} as well as a detuning
	$\Delta_i^j$ given by $\Delta_i^j=\Delta_i^{ }-\delta\omega_{12}^j=\omega_i^{
	}-\omega_{12}^j$.
	%%%%%%%%%%%%%%%%%%%%%%%%%%%%%%%%%%%%%%%%%%%%%%%%%%%%%%
	\subsection{Rate equations}
	\label{rateeq}
	%%%%%%%%%%%%%%%%%%%%%%%%%%%%%%%%%%%%%%%%%%%%%%%%%%%%%%
	A measured optical power spectrum shows the behavior of the optical power as
	a function of angular frequency. The power itself is directly related to the
	stationary photon number which again
	is determined by the system's rate equations, whose deviation is easy
	to handle in the context of the Ito formalism. We only have to calculate the
	expectation value of the QSDE for the ASE field mode product $\hat a_i^\dagger\hat
	a_{j}^{ }$ and the populations of the $j$th quantum dot, utilizing the
	characteristics of the Ito increments (see App. \ref{itoQSDE}).
	Applying the assumptions and approximations of section \ref{assumptions} and
	regarding the fact that the Ito increments commute with the system operators at
	equal time, we obtain rate equations of the QDSLD 
	\begin{equation}
	\label{rateequations}
	\begin{split}
	\dot n_i\!=\!&\sum_{j=1}^M \!\gamma G_i^j (n_i w^j \!+\! \sigma_{11}^j)
	\!-\!2\gamma^{lr}n_i \! +\! n^l\gamma^l\!+\!n^r\gamma^r
	\\
	\dot \sigma_{22}^j\!=\!& \sum_{\{k_i\}} \!\gamma G_i^j 
	(n_i w^j\!+\!\sigma_{11}^j)\!-\!\Gamma_{22}^j\sigma_{22}^j 
	+(R\!+\!\gamma_{20}^j)\sigma_{00}^j 
	\\&
	+\gamma_{21}^j (n_{21}^j\!+\!1)\sigma_{11}^j 
	\\
	\dot \sigma_{11}^j\!=\!&-\!\!\sum_{\{k_i\}} \!\gamma G_i^j
	(n_i w^j\!+\!\sigma_{11}^j)\!-\!\Gamma_{11}^j\sigma_{11}^j
	\!+\!\gamma_{21}^jn_{21}^j\sigma_{22}^j 
	\\&
	\!+\!\gamma_{10}^j(n_{10}^j\!+\!1)\sigma_{00}^j 
	\\
	\dot
	\sigma_{00}^j\!=\!&-\Gamma_{00}^j\sigma_{00}^j\!+\!\gamma_{10}^jn_{10}^j\sigma_{11}^j\!+\!R
	\sigma_{22}^j,
	\end{split}
	\end{equation}
	with mean population $\ave{\hat \sigma_{ii}^j}=\sigma_{ii}^j$ of level $\ket{i}$
	of the $j$th quantum dot
	and  detuning $\Delta_i^j$. The ``cooperativity'' strength, 
	\begin{align}
	G_i^j\equiv\frac{2}{\gamma}\frac{|g_i^j|^2\Gamma^j}{\Delta_i^{j2}+\Gamma^{j2}}
	\le \frac{2|g_i^j|^2}{\gamma\Gamma^j}, 
	\end{align}
	assesses the relative importance of the coherent atomic field coupling strength
	$g_i^j$ to the incoherent processes like pump rate $R$, external damping rates
	$\gamma^l$, $\gamma^r$ and atomic decay rates $\Gamma^j$.  
	This coupled nonlinear equation system \eqref{rateequations} can be solved
	numerically.
	However, in the following we have a closer look at two limiting cases: a single- and a  multimode radiation field interacting with a homogeneous ensemble of quantum dots, where we assume negligible occupation numbers and damping rate, $n_{10}=0$,
	$n_{21}= 0$, $\gamma_{20}= 0$, vacuum input fields, $n^r=n^l=0$, and
	equal external damping rates $\gamma^l=\gamma^r\equiv \gamma$.  
	%%%%%%%%%%%%%%%%%%%%%%%%%%%%%%%%%%%%%%%%%%%%%%%%%%%%%%%%%%%%%%%
	\subsubsection{Single-mode ASE field and identical quantum dots}
	\label{singlemode}
	%%%%%%%%%%%%%%%%%%%%%%%%%%%%%%%%%%%%%%%%%%%%%%%%%%%%%%%%%
	In case of a single-mode radiation field ($N=1$) with frequency $\omega$ acting
	on transition $\ket 1 \leftrightarrow \ket 2$ of $M$ equal quantum dots with
	transition frequency $\omega_{12}$, we find an analytical solution for the
	stationary photon number $n^s$,
	%\begin{widetext} 
	\begin{align}
	\label{D1}
	n^s&=
	a \left(1+\sqrt{1+\frac{b}{a^2}}\right),
	%  n^s=
	%  a +\sqrt{a^2+b^2} ,
	\\
	a&=\frac{\gamma_{10}R-\gamma_{21} (\gamma_{10}+R)-w_c\beta}{2w_c \gamma
		G\alpha},
	\quad
	b=\frac{\gamma_{10}R}{w_c \gamma G \alpha}, \nonumber
	\end{align}
	%\end{widetext}
	with 
	\begin{align}
	\label{alpha}
	\alpha&=3R+2\gamma_{10}, 
	\\
	\label{beta}
	\beta^{}&=(\gamma \bar G+\gamma_{21})(\gamma_{10}+2R)+\gamma_{10}R
	\end{align}
	and $\bar G=\sum_{\{k_i\}} G_i=G$. Obviously, $n^s$
	depends on the internal damping rates $\gamma_{10}$, $\gamma_{21}$ of the
	quantum dots, the incoherent pumping rate $R$ and the cooperativity strength
	$G$. 
	Here, $w_c=2/(MG)$ is the standard gain threshold, where the spontaneous
	emission term in the rate equations of Eq.~\eqref{rateequations} is neglected
	\cite{haken1984,haken1985,protsenko1999}. 
	\begin{figure}[h!]
		
		\includegraphics[width=1\columnwidth]{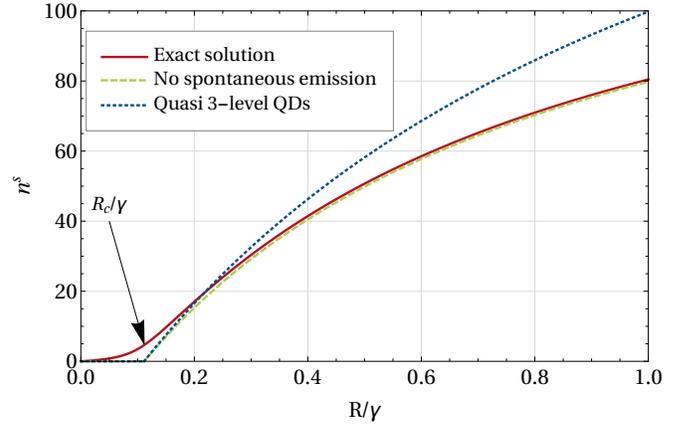}
		\caption{Stationary single-mode ($N=1$) photon number $n^s$ versus scaled
			pumping rate $R/\gamma$ with $M$=$10^3$, $\gamma_{10}=\gamma$,
			$\gamma_{21}=0.1\gamma$,  $g=\gamma$ and  $\Delta=0$. The red curve  shows
			the analytical solution \eqref{D1}, the green, dashed line represents  $n^s$ for  absent spontaneous emission. The blue, dotted line corresponds to a quasi three-level laser system of Ref.~\cite{haken1985} with
			critical pumping rate $R_c=\num{0.111}\gamma$.}
		\label{nvsR}
	\end{figure}  
	Fig.~\ref{nvsR} represents the stationary photon number $n^s$ versus scaled incoherent pumping rate $R/\gamma$ for the case of a single-mode radiation field which interacts with $M=10^3$ equal quantum dots. We chose the internal damping rates $\gamma_{21}=0.1 \gamma$, $\gamma_{10}=\gamma$, the coupling
	constant $g=\gamma$ and the detuning $\Delta=0$.
	%Here, only pumping rates smaller than the external damping cavity rate are
	%considered, $R<\gamma$, motivated by experimental measurements and theoretical
	%parameter studies \cite{Munch2009,Laucht2009,Yao2010} in terms of strong
	%coupling cavity QED.
	The red line is the analytical solution specified by Eq.~\eqref{D1}. The
	green, dashed line shows $n^s$ in absence of spontaneous emission
	processes, showing a bifurcation with critical  point, $R_c=\num{0.111}\gamma$.
	The blue, dotted curve represents the photon number of the single-mode field
	interacting with quasi-three level  quantum dots \cite{haken1984} and offers the same gain	threshold at pumping rate $R_c$ as the green line. The
	comparison between the red and green, dashed curve points out, that in case of the red
	line the threshold is smeared out and the characteristic amplified spontaneous emission
	behavior is observable 
	\cite{Peters1971c,Ning_whatis,Blazek2012}. 
	The blue line highlights an increasing deviation from the red and green line for higher values of   $R$.
	
\begin{figure}[h!]
		
		\includegraphics[width=1\columnwidth]{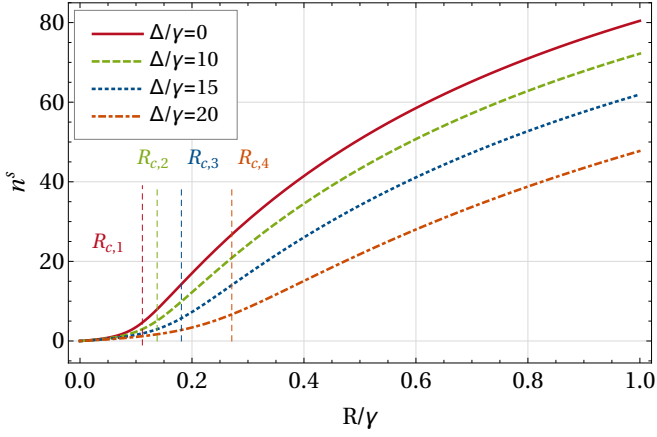}
		%    \caption{Stationary single-mode ($N=1$) photon number $n^s$ versus scaled
		%pumping rate $R/\gamma$ for a fix coupling constant $g=\gamma$ and parameters
		%$M=$\num{1000}, $\gamma_{21}=0.1\gamma$, $\gamma_{10}=\gamma$. The detuning has
		%values $\Delta=0$ (red line; critical pumping rate
		%$R_{c,1}=$\num{0.111}$\gamma$), $\Delta=10\gamma$ (green, dashed line; critical
		%pumping rate $R_{c,2}=$\num{0.138}$\gamma$), $\Delta=15\gamma$ (blue, dotted
		%line; critical pumping rate $R_{c,3}=$\num{0.181}$\gamma$) and
		%$\Delta=20\gamma$ (orange, dashed-dotted line; critical pumping rate
		%$R_{c,4}=$\num{0.271}$\gamma$) with the corresponding critical pumping rates
		%$R_c$.  }
		\caption{Stationary single-mode ($N=1$) photon number $n^s$ versus scaled
			pumping rate $R/\gamma$ with parameters $g=\gamma$, $M=10^3$,
			$\gamma_{21}=0.1\gamma$, $\gamma_{10}=\gamma$ and different detuning $\Delta=0$
			(red line; $R_{c,1}=\num{0.111}\gamma$), $\Delta=10\gamma$ (green, dashed line; 
			$R_{c,2}=\num{0.138}\gamma$), $\Delta=15\gamma$ (blue, dotted line; 
			$R_{c,3}=\num{0.181}\gamma$) and $\Delta=20\gamma$ (orange, dashed-dotted line;
			$R_{c,4}=\num{0.271}\gamma$) as well as corresponding critical pumping rates
			$R_c$.  }
		\label{nvsRdiffdetuning}
	\end{figure}

%	Regarding  the stationary photon number as a function of the pumping rate for
%	increasing detuning, the gain threshold $R_{c,k}$ ($k=1,2,3,4$) tends to higher
%	values of $R$. Fig.~\ref{nvsRdiffdetuning} reflects this behavior for fix
%	coupling constant $g=\gamma$, number of quantum dots $M=10^3$ and decay rates
%	$\gamma_{21}=0.1\gamma$ and $\gamma_{10}=\gamma$. 
For increasing detuning $\Delta$, the stationary photon number in terms of pumping rate $R$ offers an
	increasing gain threshold $R_{c,k}$ ($k=1,2,3,4$) as demonstrated in Fig.~\ref{nvsRdiffdetuning} with coupling constant $g=\gamma$, number of quantum dots $M=10^3$ and decay rates
	$\gamma_{21}=0.1\gamma, \gamma_{10}=\gamma$. 
	The detuning is given by
	$\Delta=0$ (red line; $R_{c,1}=\num{0.111}\gamma$), $\Delta=10\gamma$ (green,
	dashed line;  $R_{c,2}=\num{0.138}\gamma$), $\Delta=15\gamma$ (blue, dotted
	line;  $R_{c,3}=\num{0.181}\gamma$) and $\Delta=20\gamma$ (orange, dashed-dotted
	line;  $R_{c,4}=\num{0.271}\gamma$). 
	%The strongest rise of the photon number in the regime $0<R<\gamma$ is
	%observable for zero detuning.
	For decreasing detuning $\Delta$, the slope of the function $n^s(R/\gamma)$ increases until the saturation regime is reached.
	
\begin{figure}[h!]
		\includegraphics[width=1\columnwidth]{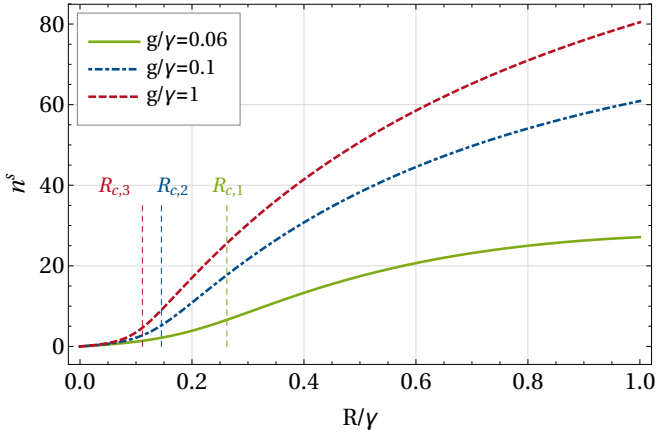}
		\caption{Stationary single-mode ($N=1$) photon number $n^s$ versus scaled
			pumping rate $R/\gamma$ for vanishing detuning, i.e. $\Delta=0$, parameters
			$M=10^3$, $\gamma_{21}=0.1\gamma$, $\gamma_{10}=\gamma$ and varying coupling
			constant $g=\num{0.06}\gamma$ (green line;  $R_{c,1}=\num{0.262}\gamma$),
			$g=0.1\gamma$ (blue, dotted-dashed line;  $R_{c,2}=\num{0.145}\gamma$) and
			$g=\gamma$  (red, dashed line;  $R_{c,3}=\num{0.111}\gamma$).}
		\label{nvsRdiffg}
	\end{figure}

	A similar behavior is observable for $n^s(R/\gamma)$ for decreasing coupling $g$
	as shown in Fig.~\ref{nvsRdiffg} with decay rates $\gamma_{21}=0.1\gamma$,
	$\gamma_{10}=\gamma$, detuning $\Delta=0$ and quantum dot number $M=10^3$. The coupling constants are chosen to be 
	$g=0.06\gamma$ (green line;  $R_{c,1}=\num{0.262}\gamma$) and $g=0.1\gamma$
	(blue, dotted-dashed line; $R_{c,2}=\num{0.145}\gamma$), $g=\gamma$ (red, dashed
	line;  $R_{c,3}=\num{0.111}\gamma$). For decreasing coupling constant $g$, the
	gain threshold tends to higher values of pumping rate $R$. 
	%In the strong coupling regime, where $g \gg \gamma$, the photon number for
	%varying pumping rates follows the same behavior as in the case of $g=\gamma$
	%corresponding to the red, dashed curve in Fig.~\ref{nvsRdiffg}. Thus, within
	%this microscopic model, the photon number exhibits a kind of "cut off"  at
	%$g=\gamma$. 
	%%%%%%%%%%%%%%%%%%%%%%%%%%%%%%%%%%%%%%%%%%%%%%%%%%%%%%%%%%%%%%
	\subsubsection{Multimode ASE field with identical quantum dots}
	\label{multimode}
	%%%%%%%%%%%%%%%%%%%%%%%%%%%%%%%%%%%%%%%%%%%%%%%%%%%%%%%
	For a multimode radiation interacting with equal quantum dots inside the diode
	system, the rate equations \eqref{rateequations} reduce to
	\begin{equation}
	\label{rateequationEqualDots}
	\begin{split}
	\dot n_i\!=&M \gamma G_i (n_i w+\sigma_{11})
	-2\gamma^{lr}n_i  + n^l\gamma^l+n^r\gamma^r
	\\
	\dot \sigma_{22}\!=& \sum_{\{k_i\}} \!\gamma G_i 
	(n_i w+\sigma_{11})-\Gamma_{22}\sigma_{22} +(R+\gamma_{20})\sigma_{00}
	\\&
	+\gamma_{21} (n_{21}+1)\sigma_{11} 
	\\
	\dot \sigma_{11}\!=&-\sum_{\{k_i\}} \gamma G_i
	(n_i w+\sigma_{11})+\gamma_{21}n_{21}\sigma_{22}-\Gamma_{11}\sigma_{11}
	\\&  
	+\gamma_{10}(n_{10}+1)\sigma_{00}
	\\
	\dot \sigma_{00}\!=&-\Gamma_{00}\sigma_{00}+\gamma_{10}n_{10}\sigma_{11}+R
	\sigma_{22}.
	\end{split}
	\end{equation}
	
	To find an analytical solution of the stationary photon number $n^s$, we
	introduce an order parameter,
	\begin{align}
	\label{D2}
	\varphi \equiv \gamma\sum_{\{k_i\}} G_i n_i^s, \quad n^s=\sum_{\{k_i\}} n_i^s,
	\end{align}
	for the diode system  in detailed balance \cite{Mcneil1975}. The order parameter
	corresponds to a weighted stationary total photon number, $n^s$. The stationary
	photon number of mode  $i$,
	\begin{align}
	\label{D3}
	n_i^s(\varphi)=\frac{\sigma_{11}^s(\varphi)}{w_i^c-w^s(\varphi)},
	\quad w_i^c=\frac{2}{MG_i},
	\end{align}
	is a function of $\varphi$ with stationary excited state and inversion,
	\begin{align}
	\sigma_{11}^s(\varphi)&=\frac{\varphi(\gamma_{10}+R)+\gamma_{10}R}{\alpha\varphi+\beta},\\
	w^s(\varphi)&=\frac{\gamma_{10}R-(\gamma\bar
		G+\gamma_{21})(\gamma_{10}+R)}{\alpha\varphi+\beta}.
	\end{align}
	$\alpha$ and $\beta$ depend on the incoherent pumping rate $R$ (see
	Def.~\eqref{alpha} and \eqref{beta}) and  $\bar G=\sum_{\{k_i\}} G_i$ is the sum
	over all $N$ cooperativity strength, $G_i$. 
	Inserting Eq.~\eqref{D3} in Eq.~\eqref{D2}, we get a closed relation for  $\varphi$,
	\begin{align}
	\label{phi}
	\varphi
	=\sum_{\{k_i\}}\frac{\gamma G_i\sigma_{11}^s(\varphi)}{w_i^c-w^s(\varphi)} 
	=\frac{\sigma_{11}^s(\varphi)}{w^s(\varphi)}
	\Bigl[
	\sum_{\{k_i\}}\frac{\gamma G_i w_i^c}{w_i^c-w^s(\varphi)}-\gamma\bar G
	\Bigr],
	\end{align}
	whose  solution and therefore the photon number itself can be calculated numerically.
	
	In good approximation, we replace $w_i^c$ in the enumerator of the second term
	in the brackets of Eq.~\eqref{phi} by the minimum gain threshold
	$w^c_{\text{min}}=2/(MG_\text{max})$ with $G_\text{max}\equiv G(g_i= 2/3
	g_\text{max},R)$ and maximum coupling constant $g_\text{max}$.  Eq.~\eqref{phi} simplifies to 
	\begin{align}
	\label{approxphi}
	\varphi&\approx -\gamma\bar
	G\frac{\sigma_{11}^s(\varphi)}{w^s(\varphi)}+w^c_{\text{min}}\frac{\varphi}{w^s(\varphi)}\nonumber
	\\
	&=
	\frac{\gamma_{10}R-\gamma_{21}(\gamma_{10}+R)-w_\text{min}^c\beta}{2w_\text{min}^c
		\alpha}\\
	&+ \sqrt{\left(
		\frac{\gamma_{10}R-\gamma_{21}(\gamma_{10}+R)-w_\text{min}^c\beta}{2w_\text{min}^c
			\alpha}\right)^2+\frac{\gamma \bar G \gamma_{10}R}{w_\text{min}^c\alpha}}
	\nonumber
	\end{align}
	and we can specify an approximate solution of the stationary intrawaveguide photon number $n^s$ (see Eq.~\eqref{D3}) of a  QDSLD based on our microscopic theory. 
	
\begin{figure}[h!]
		\includegraphics[width=1\columnwidth]{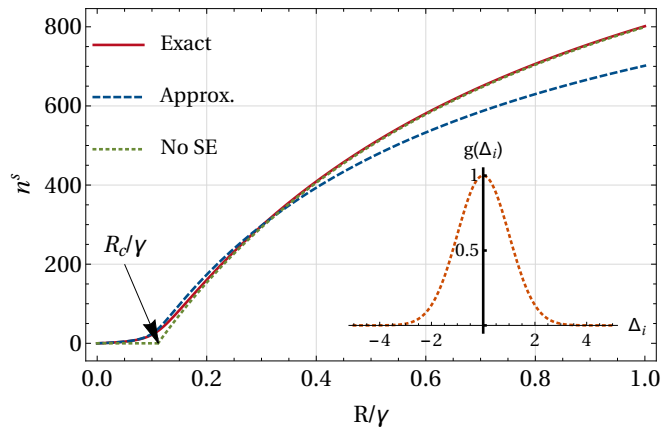}
		\caption{Photon number $n^s$ versus scaled incoherent pumping rate $R/\gamma$ 
			for a  $N=30$ mode radiation field, interacting with $M=10^4$ equal quantum dots. The coupling constant $g_i$ is Gaussian-shaped (see Eq.~\eqref{gauss}) with $g_0=\gamma$, $\bar\Delta=0$ and  $\sigma=\gamma$.
			Decay rates and detuning are $\gamma_{21}=0.1\gamma$,
			$\gamma_{10}=\gamma$ and $\Delta_i= i\gamma $. The red line reflects the 
			numerical photon number, the blue, dashed line is the
			approximated solution described in Eq.~\eqref{approxphi} and 
			the green, dotted line is the solution for absent  spontaneous emission (SE)
			processes \cite{haken1984} with critical
			pumping rate $R_c=\num{0.111}\gamma$.}
		\label{multi_nvsR_noSE_approx_exact}
	\end{figure}

	\paragraph{Rectangular shape of the cooperativity strength} 
	In case of a QDSLD with identical QDs and rectangular distributed cooperativity
	strength $G_i=G$, the resulting rate equations agree with the rate equations of
	a single-mode QDSLD leading to an exact analytical solution for the stationary
	photon number $n^s=\sum_{\{k_i\}}n_i^s$ given by Eq.~\eqref{D1}.
	\paragraph{Gaussian shaped coupling} 
	Motivated by the observed Gaussian-shaped spectrum (see Fig.~\ref{spectrum}),
	Fig.~\ref{multi_nvsR_noSE_approx_exact} visualizes  the stationary photon number
	$n^s$ as a function of $R/\gamma$ for a radiation field composed of $N=30$ modes which interacts with
	$M=10^4$ equal quantum dots with Gaussian coupling constant, 
	\begin{align}
	\label{gauss}
	g(\Delta_i)=g_0~\text{exp}\left(-\frac{(\Delta_i - \bar \Delta)^2}{2
		\sigma^2}\right),
	\end{align}
	where we have chosen an amplitude $g_0=\gamma$, expectation value $\bar\Delta=0$
	and width $\sigma=\gamma$. 
	Damping rates and detuning are  $\gamma_{21}=0.1\gamma$,
	$\gamma_{10}=\gamma$ and $\Delta_i=  i\gamma $. The red line shows the exact,
	numerically calculated solution of $n^s$ and the blue, dashed line is the
	approximated one, specified by Eq.~\eqref{D2}, \eqref{D3} and 
	\eqref{approxphi}. The green, dotted line is the solution of $n^s$ for
	negligible spontaneous emission contribution in the sense of
	Ref.~\cite{haken1984} with critical pumping rate, $R_c=\num{0.111}\gamma$.
	Comparison of these three lines points out that the approximate photon number (blue, dashed)
	and the exact (red) solution agrees well  for smaller values of $R$. For $R>0.4\gamma$, their
	deviation  increases, in contrast to the case of absent spontaneous emission processes approaching the exact
	solution.
%	Furthermore, depending on the values of the Gaussian distributed coupling
%	constants $g_i$ of the multimode radiation field, the deviation between the
%	multimode and single-mode photon number (black, dashed-dotted line) as a
%	function of the pumping rate $R$ is more or less significant. Here, the photon
%	number of the single-mode radiation field for increasing $R/\gamma>0.4$ becomes
%	lager than in the multimode case.
	
\begin{figure}[h!]
		\includegraphics[width=.9\columnwidth]{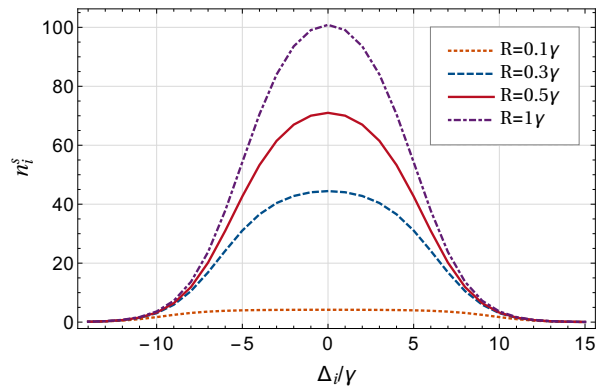}
		\caption{Stationary multimode photon number $n^s$ of the radiation field
			with $N=30$ modes versus scaled detuning $\Delta_i/\gamma$ for different scaled
			pumping rates $R/\gamma=0.1,0.3,0.5,1$. Here, the coupling constants $g_i$ are
			Gaussian distributed (see Eq.~\eqref{gauss}) with mean value
			$\bar\Delta/\gamma=N/10$ and standard deviation $\sigma/\gamma=N/5$. The damping
			rates are chosen to be $\gamma_{10}=\gamma$, $\gamma_{21}=0.1\gamma$ and the
			gain medium consists of $M=10^4$ quantum dots.}
		\label{nvsomega}
	\end{figure}

	Fig.~\ref{nvsomega} shows the intrawaveguide photon number $n_i^s$ of a
	radiation field with $N=30$ modes as a function of detuning of the $i$th mode
	with Gaussian distributed coupling constant, described by Eq.~\eqref{gauss},
	having an amplitude $g_0=\gamma$, a mean value $\bar\Delta/\gamma=N/10$ and
	a standard deviation $\sigma/\gamma=N/5$. The external damping rates are
	again $\gamma_{21}=0.1\gamma$ and $\gamma_{10}=\gamma$. The scaled pumping rate
	$R/\gamma$ was chosen to be $R/\gamma=0.1,0.3,0.5,1$. The gain medium consists
	of $M=10^4$ quantum dots. This photon number versus frequency is of
	Gaussian-like shape  with nonlinear corrections and exhibits an increasing
	central frequency and width when increasing the pumping rate~$R$. Thus, the
	photon number is coined by the shape of $G_i$. 
	%%%%%%%%%%%%%%%%%%%%%%%%%%%%%%%%%%%%%%%%%%%%%%%%%%%%%%%%%%%%%%%
	\section{Output spectrum}
	\label{output}
\begin{figure}[h!]
		\centering
		\includegraphics[width=.8\columnwidth]{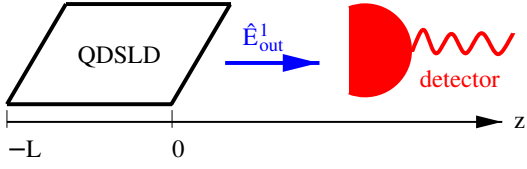}
		\caption{Sketch of the detection process to determine the first-order
			correlation function of the emitted QDSLD light at position $\vec r_\text{d}$
			using a detector with cross-section area $A_d$.}
		\label{detection}
	\end{figure}

	Up to now, we have studied only the intrawaveguide system in terms of rate
	equations, Eq.~\eqref{rateequations}. To compare our theoretical predictions
	with the experimental data, we study the optical power spectrum of a QDSLD. The
	physical problem is sketched in Fig.~\ref{detection}.
	
	 By the help of a single
	detector at fixed position $\vec r_d=(x_d,y_d,z_d)$, the first-order correlation
	function is determined. 
	%\begin{figure}[h!]
	%  \centering
	%    \scalebox{.8}{\input{pictures/detection/scetchdetectornew.pdf_t}}
	%    \caption{Sketch of the detection process to determine the first-order
	%correlation function of the emitted QDSLD light at position $\vec r_\text{d}$
	%using a detector with cross-section area $A_d$.}
	%  \label{detection}
	%\end{figure}
	In App~\ref{emptyCavity}, we investigate in more detail the spectral density of
	an empty diode, i.e. without any gain medium. Here, we follow the main ideas and
	proceedings of App. B and apply them to the case of a truly incoherent QDSLD.
	
	The real output temporal field correlation function at the position of the
	detector with cross-section area $A_d$,
	\begin{align}
	G^{(1)}(\tau)=\lim_{t\to \infty}\int\limits_{A_d}
	dx_ddy_d\ave{\hat{\vec{E}}^{1{(-)}}_\text{out}(\vec
		r_d,t)\hat{\vec{E}}^{1(+)}_\text{out}(\vec r_d,t+\tau)},
	\end{align}
	is directly connected to the power spectrum according to the Wiener-Khintchine
	theorem, 
	%\cite{louisell1973},
	\begin{align}
	\label{svsom}
	S(\omega)=\frac{\mathcal C}{\pi}\,\text{Re}\int_{0}^\infty
	d\tau~e^{\ii\omega\tau}
	G^{(1)}(\tau),
	\quad\mathcal C=2\epsilon_0 c .
	\end{align}
	Relying on the concept of the input-output formalism (see App.
	\ref{emptyCavity}) the output operator of channel $\alpha$, 
	\begin{align}
	\label{inoutrel}
	d\hat b_{i,\text{out}}^{\alpha}(t)&
	=d\hat b_{i,\text{in}}^{\alpha}(t)+\zeta_i^\alpha \hat a_i(t)dt
	\end{align}
	with $ \zeta_i^\alpha$ defined in \eqref{zeta},
	depends on the input operators of channel $\alpha$ and the internal modes $\hat
	a_i$ of the quantized field. According to the orientation of the output facets
	shown in Fig.~\ref{sld}, channel $1$ enters the diode on the right hand side at
	$z=0$. The electric output field $\hat{\vec E}_\text{out}^1$ is assumed to be
	guided in a single-mode fiber yielding to a radiation field propagating along
	the $z$-direction. It strikes the detector at $z=z_d$.
	
	Evaluating the two-time correlation of the output operators with $t<t'=t+\tau$
	in consideration of Eq.~\eqref{inoutrel}, we find
	\begin{align}
	\label{cor}
	\begin{split}
	\ave{&d\hat b_{i,\text{out}}^{1\dagger}(t) d\hat b_{j,\text{out}}^{1^{ }}(t')}=
	\ave{d \hat{b}_{i,\text{in}}^{1\dagger}(t)d \hat{b}_{j,\text{in}}^{1}(t')} \\
	&
	+\zeta_i^{1\ast} \ave{\hat{a}_i^\dagger(t)d \hat{b}_{j,\text{in}}^{1}(t')}dt
	+\zeta_j^{1}\ave{d \hat{b}_{i,\text{in}}^{1\dagger}(t) \hat a_{j} (t') }dt' \\
	&+\zeta_i^{1\ast}\zeta_j^{1}\ave{\hat a_i^\dagger (t)\hat a_{j}^{ } (t')}dtdt'.
	\end{split}
	\end{align}
	In the following, we choose vacuum as an input, where the first three terms in
	Eq.~\eqref{cor} vanish. Thus, the output field is directly related to the
	first-order autocorrelation function of the ASE field modes at different time
	events $t<t'$,
	\begin{align}
	\begin{split}
	\label{4.1}
	d\ave{\hat a_i^{\dagger}(t)\hat a_{s}^{ }(t')}=&
	-\ii(\Delta_{s}-\ii\gamma^{lr}\bigr)\ave{\hat a_i^{\dagger}(t) \hat a_{s}^{
		}(t')}dt' 
	\\
	&
	+\sum_{j=1}^M g_{s}^{j\ast}\ave{\hat a_i^\dagger (t)\hat \sigma_{21}^j(t')}dt'.
	\end{split}
	\end{align}
	Applying again the decorrelation approximation to $\ave{\hat a_i^\dagger (t)\hat
		\sigma_{21}^j(t')}$ (see Eq.~\eqref{decorr}) and adiabatic elimination,
	Eq.~\eqref{4.1} reduces to
	%\begin{align}
	%\label{4.2}
	%  d&\ave{\hat a_i^{\dagger}(t) \hat a_{s}^{ }(t')}\nonumber \\
	%&  =-\Bigl(\ii\Delta_{s}+\gamma^{lr}-\sum_{j=1}^M \frac{|g_{s}^j|^2 
	%w^j(t')}{\Gamma_{21}^j+\ii\delta\omega_{12}^j}\Bigr)
	%  \ave{\hat a_i^{\dagger}(t) \hat a_{s}^{ }(t')}dt' \nonumber\\
	%&+\sum_{j=1}^M \sum_{\{k_m\}\atop m\neq s}\frac{ g_m^jg_{s}^{j\ast} 
	%w^j(t')}{\Gamma_{21}^j+\ii\delta\omega_{12}^j} \times \ave{\hat
	%a_i^{\dagger}(t) \hat a_{m}^{ }(t')}dt'.
	%\end{align}
	\begin{align}
	\label{4.2}
	\begin{split}
	d\ave{\hat a_i^{\dagger}(t) &\hat a_{s}^{ }(t')}
	=-\Bigl(\eta_s(t')+\ii \chi_s(t')\Bigr)
	\ave{\hat a_i^{\dagger}(t) \hat a_{s}^{ }(t')}dt' 
	\\
	&+\sum_{j=1}^M \sum_{\{k_m\}\atop m\neq s}\frac{ g_m^jg_{s}^{j\ast} 
		w^j(t')}{\Gamma_{21}^j+\ii\delta\omega_{12}^j} \times \ave{\hat a_i^{\dagger}(t)
		\hat a_{m}^{ }(t')}dt'
	\end{split}
	\end{align}
	with 
	%$
	%\eta_s(t')= \gamma^{lr}
	%-\sum_{j=1}^M\nicefrac{|g_s^j|^2w^{j}(t')
	%\Gamma_{21}^j}{\Gamma_{21}^{j2}+\delta \omega_{12}^{j2}} $ and 
	%$\chi_s(t')=\Delta_s+\sum_{j=1}^M \nicefrac{|g_s^j|^2 \delta \omega_{12}^{j}
	%w^{j}(t')}{\Gamma_{21}^{j2}+\delta \omega_{12}^{j2}} $.
	\begin{align}
	\chi_i(t)-\ii \eta_i(t)=\Delta_i-\ii \gamma^{lr}+\sum_{j=1}^M \frac{|g_i^j|^2
		w^j(t)}{\delta\omega_{12}^j-\ii \Gamma_{21}^j}.
	\end{align}
	We claim that off-diagonal elements of the coupling matrix are
	small in  comparison to the diagonal one. Furthermore, in first-order perturbation theory we disregard the last term of Eq.~\eqref{4.2}. 
	
	In equilibrium, the inversion tends to a constant value, 
	$\lim_{t\to \infty}\ave{\hat w^j(t+\tau)}=w^{js}=\text{const}.$. Clearly, as 
	we want to investigate the stationary power spectrum of a QDSLD \eqref{svsom}, the
	time dependent inversion arising in the expressions of $\eta_s(t')$ and
	$\chi_s(t')$ of Eq.~\eqref{4.2} can be replaced by $w^{js}$
	% and the field correlation function with $t'>t$ reads
	%\begin{align}
	% \ave{\hat a_i^\dagger(t)\hat a_{j}^{
	%}(t')}=n_i(t)e^{-(\ii\chi_i+\eta_i)(t'-t)}\delta_{ij}.
	%\end{align}
	%Here, the propagation and absorption coefficients \cite{saleh2007}, $\chi_i$
	%and $\eta_i$, follow from Taylor expansion for small QD transition deviation
	%$\delta \omega_{12}^j$ with
	%\begin{align}
	%\chi_i&=\Delta_i+\sum_{j=1}^M \frac{|g_i^j|^2w^{js}}{\Gamma_{21}^{j2}+\delta
	%\omega_{12}^{j2}}\delta \omega_{12}^{j} 
	%\approx \Delta_i,
	%\\
	%\eta_i&=
	% \gamma^{lr}
	%-\sum_{j=1}^M\frac{|g_i^j|^2w^{js} \Gamma_{21}^j}{\Gamma_{21}^{j2}+\delta
	%\omega_{12}^{j2}} 
	%\approx \frac{\gamma^l+\gamma^r-\xi_i}{2}
	%\end{align}
	%with $\xi_i=2 \sum_{j=1}^M\nicefrac{|g_i^j|^2w^{js}}{\Gamma_{21}^{j}}$.
	. For small QD transition deviations $\delta \omega_{12}^j$, we can replace
	propagation and absorption coefficients \cite{saleh2007} according to 
	\begin{align}
	\chi_i\approx \Delta_i, \quad \eta_i\approx \gamma^{lr}-\xi_i,\quad  \xi_i=2
	\sum_{j=1}^M\frac{|g_i^j|^2w^{js}}{\Gamma_{21}^{j}},
	\end{align}
	and the field correlation function with $t'>t$ reads
	\begin{align}
	\ave{\hat a_i^\dagger(t)\hat a_{j}^{ }(t')}=n_i(t)e^{-\ii(\chi_i-\ii
		\eta_i)(t'-t)}\delta_{ij}.
	\end{align}
	
	Finally, the correlation function for the output operators $d\hat
	b^{1}_{i,\text{out}}$ at different time events $t$, $t'$ reads
	\begin{align}
	\ave{d\hat b_{i,\text{out}}^{1\dagger}(t) & d\hat b_{j,\text{out}}^{1}(t')}=
	% \frac{\gamma^r}{2} n_i(t)e^{-\ii(\chi_i-\ii \eta_i)(t'-t)}\delta_{ij}dtdt' .
	\frac{\gamma^r}{2}\ave{\hat a_i^\dagger(t)\hat a_{j}^{ }(t')}\delta_{ij}dtdt'
	.
	\end{align}
	The stationary optical power spectrum in the original Heisenberg picture
	\eqref{Heisenberg} is given by
	\begin{align}
	\label{spectrumdis}
	S(\omega)=\sum_{\{k_i>0\}} P_i \mathcal{L}_{\Gamma_i}(\omega-\omega_i) \, n_i^s
	%\approx P \sum_{\{k_i>0\}}  \mathcal{L}_{\Gamma_i}(\omega-\omega_i) n_i^s,
	\quad
	P_i=\frac{\hbar \omega_i c\gamma^r}{L\Delta\omega_i^r}
	\end{align}
	with optical power $ P_i$ and Lorentzian function
	\begin{align*}
	\mathcal L_{\Gamma}(\omega)=
	\frac{1}{\pi}\frac{\Gamma}{(\Gamma/2)^2+\omega^2},
	\quad\Gamma=2(\gamma^{lr}-\xi),
	\end{align*}
	with  $\int_{-\infty}^\infty d\omega \mathcal L_\Gamma(\omega)=1$.
	% with the stationary intrawaveguide photon number $n_i^s$.  In the last step of
	%\eqref{spectrumdis}, we assumed that the frequencies $\omega_i$ show negligible
	%small deviation from the central frequency $\bar \omega$ and the corresponding
	%frequency separation between adjacent modes are equivalent, i.e.
	%$\Delta\omega_i^r=\Delta\omega^r=(\omega_N-\omega_1)/(N-1)$.
	%The sum in \eqref{spectrumdis} can be approximated by the first term of the
	%Euler-MacLaurin series, i.e. the sum is replaced by the integral,
	%$\sum_{\{k_i\}}  f(\omega_i)\rightarrow 1/\Delta\omega^r
	%\int_{\omega_1}^{\omega_N}f(\omega)$. The continuous power spectrum  reads
	Assuming negligible small deviation of $\omega_i$  from a central frequency
	$\bar \omega$  as well as equidistant frequency separations, i.e.
	$\Delta\omega_i^r=\Delta\omega^r$, the continuous optical power spectrum
	\begin{align}
	\label{specconti}
	S(\omega)&=
	\frac{P}{\Delta\omega^{r}}\int_{-\infty}^{\infty} d\omega'
	\mathcal{L}_\Gamma(\omega-\omega') n^s(\omega') , 
	\quad
	P=\frac{\hbar \bar\omega c\gamma^r}{L\Delta\omega^r}
	\end{align}
	%with 
	%\begin{align*}
	%\mathcal{L}(\omega-\omega')&=
	%\frac{1}{2\pi}
	%\frac{\gamma^l+\gamma^r}{(\frac{\gamma^l+\gamma^r}{2})^2+(\omega-\omega')^2}.
	%\end{align*}
	is a convolution of a Lorentzian curve with the stationary intrawaveguide photon
	number specified by the rate equations \eqref{rateequations}.

\begin{figure}[h!]
		\includegraphics[width=1\columnwidth]{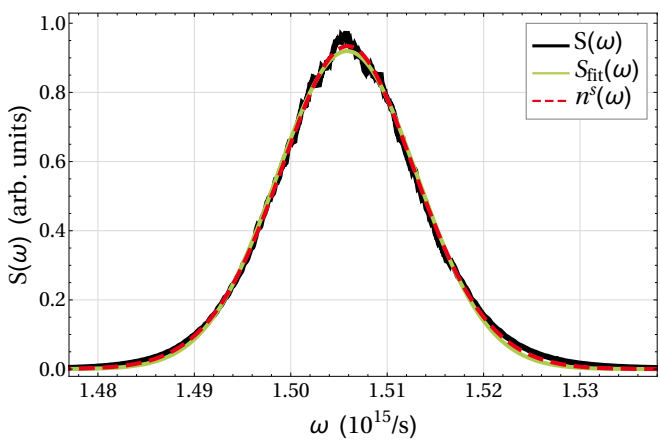}
		\caption{Experimental power spectrum $S(\omega)$ (black) and Gaussian
			fit (green line) (see Fig.~\ref{spectrum})  and stationary intrawaveguide
			photon number $n^s(\omega)$ (red, dashed line) for $M=10^4$ identical QDs and Gaussian shaped $G(\omega)$ (see
			Eq.~\eqref{cooperativity}) with fit parameters listed in Tab.~\ref{tablefit}.}
		\label{datamodel}
	\end{figure}

	Fig.~\ref{datamodel} visualizes the spectral density as a function of angular
	frequency. The black dots are the experimental data and the green line is a
	Gaussian fit, both already depicted in Fig.~\ref{spectrum}.  
	The red, dashed curve reflects the solution of our quantum theory of
	light-emitting QDSLDs. More precisely,  it shows the intrawaveguide photon
	number in case of $M=10^4$ equal quantum dots and Gaussian shaped cooperativity
	strength (see Eq.~\eqref{cooperativity}).  
	Values of pumping rate and internal damping rates are again chosen to be
	$R=0.5\gamma$, $\gamma_{10}=\gamma$ and $\gamma_{21}=0.1\gamma$.
	%The corresponding fit parameters listed in Tab.~\ref{tablefit}, are numerically
	%determined by finding the minimum of the norm of the differences of the
	%successive differences of all logarithmic photon numbers and the corresponding
	%successive differences of all logarithmic experimental data, weighted by the
	%data $S(\omega_i)$ itself. 
	Practically, the numerical solution of $n^s(\omega)$ is investigated by solving
	the rate equations and using Newton's method. 
	Both, Gaussian fit and numerical solution of the internal photon number fit the
	data remarkably well revealing a power spectrum with a THz spectral width. Thus, the
	optical spectrum of a QDSLD is primarily affected by the stationary intrawaveguide
	photon number itself. 
	%The Gaussian-like behavior of $n^s(\omega)$ suggests that the shape of
	%$S(\omega)$ is mainly specified by the intrawaveguide photon number itself.
	%However, this shape of the number of photons originates from the choice of the
	%cooperativity factor $G(\omega)$.
	Its shape again depends significantly on the distribution of the cooperativity
	strength $G(\omega)$.
	%At first sight, both lines fit the data very well, but the red line reflects
	%the measured spectral density slightly better, especially at the outer region. 
	%\begin{figure}[h!]
	%    \includegraphics[width=1\columnwidth]{pictures/logplotmodel.pdf}
	%    \caption{Change of the power spectrum (see Fig.~\ref{datamodel}) weighted
	%with the experimental data $S(\omega_i)$. }
	%    \label{datamodellog}
	%\end{figure}    
	%This suspicion is substantiated by Fig.~\ref{datamodellog}, showing the
	%differences of the logarithmic spectrum weighted with the experimental data
	%(black dots). The antisymmetric behavior is observable for the theoretical
	%model of the intrawaveguide photon number as well as in case of an optimal
	%fitted Gaussian. Thus, the shape of the intrawaveguide photon number and
	%therefore the power spectral density itself, is formed directly by the shape of
	%the cooperativity factor. 
	Knowing all experimental parameters, we are able to replicate each measured
	optical power spectrum of a QDSLD from microscopic considerations. Furthermore,
	knowing the experimental data, we can specify unknown, relevant parameters like
	coupling strength between QD and ASE field and so on.
	%%%%%%%%%%%%%%%%%%%%%%%%%%%%%%%%%%%%%%%%%%%%%%%%%%%%%%%%%%%%%%%
	\section{Conclusion}
	\label{conclusion}
	%%%%%%%%%%%%%%%%%%%%%%%%%%%%%%%%%%%%%%%%%%%%%%%%%%%%%%%%%%%%%%%%%%%%%%%%%%%%%%
	In this article, we have studied the amplified spontaneous emission of
	light-emitting quantum dot superluminescent diodes in terms of first-order
	temporal autocorrelation function. 
	We presented a microscopic theory of QDSLDs, which considers the specific gain
	medium, formed by an inhomogeneous ensemble of QDs, as well as tilted end
	facets. The former was modeled by distinguishable three-level systems,
	incoherently pumped and interacting with a multimode radiation field, which
	couples to the outside by two beam splitters enclosing the active medium. 
	We derived quantum stochastic differential equations in the quantum white noise
	limit, which allow to calculate first- as well as higher-order temporal
	correlation functions.
	
	The resulting rate equations for the optical power densities and level
	occupations of the quantum dots within the QDSLD were studied for the limiting
	cases of a single-mode and a multimode ASE radiation field interacting with $M$
	equal quantum dots. 
	We showed, that the  stationary intrawaveguide photon number as a function of
	the incoherent pumping rate agrees well with experimental results, where the
	typical amplified spontaneous emission transition is visible. 
	
	We have applied the input-output formalism to determine the optical power
	spectrum at the detector's position. As a main result, we showed that this
	spectral density is a convolution of a Lorentzian curve with the intrawaveguide
	photon number, whose shape is coined by the shape of the cooperativity strength.
	%%%%%%%%%%%%%%%%%%%%%%%%%%%%%%%%%%%%%%%%%%%%%%%%%%%%%%%%%%%%%%%%%%%%%%%%%%%%%%%
	%%%%%%%%%%%%%%%%%%%%%%%%%%%%%%%%%%%%%%%%%%%%%%%%%%%%%%%%%%%%%%%%%%%%%%%%%%%%%%%
	\section{Acknowledgment}
	\label{acknowledgment}
	We like to thank S\'{e}bastien Blumenstein (name of birth Hartmann) for the
	experimental data, all experimental details and fruitful discussions.
	Furthermore, we thank the German Aerospace Center (DLR) (50 WM 1557) and the
	Deutsche Forschungsgemeinschaft (DLR) for financial support.
	%%% %%%%%%%%%%%%%%%%%%%%%%%%%%%%%%%%%%%%%%%%%%%%%%%%%
	%%%%%%%%%%%%%%%%%%%%%%%%%%%%%%%%%%%%%%%%%%%%%%%%
	\appendix
	%%%%%%%%%%%%%%%%%%%%%%%%%%%%%%%%%%%%%%%%%%%%%%%%%%%%%%%%%%%%%%%
	\section{Quantum Ito calculus}
	\label{itoQSDE}
	One can analyze the dynamics of Heisenberg operators $\hat Y(t)$  and 
	$\{\hat X_m(t)\}$ of a small open quantum system coupled to $\mathcal M$
	reservoirs
	by the quantum stochastic differential equation (QSDE)  
	\begin{equation}
	\label{qsde}
	\begin{split}
	d\hat Y=
	&-\frac{i}{\hbar} \comut{\hat Y}{\hat H_s}dt 
	+\sum_{m=1}^{\mathcal M}\left \{ \right .\\
	&\gamma_m 
	\bigl(
	X_m^\dagger\comut{\hat Y}{ \hat X_m}
	-\comut{\hat Y}{\hat X_m^\dagger} \hat X_m
	\bigr)\bigl(n_m+1\bigr)dt \\
	+&\gamma_m \bigl(
	-\comut{\hat Y}{\hat X_m}\hat X_m^\dagger
	+\hat X_m \comut{\hat Y}{\hat X_m^\dagger}\bigr) n_m dt  \\
	+&\left . d\hat B^\dagger_m(t)\comut{\hat Y}{\hat X_m}
	-\comut{\hat Y}{\hat X_m^\dagger}d\hat B_m(t)
	\right \},
	\end{split}
	\end{equation}
	in the Ito interpretation \cite{Gardiner1985,quantumnoise}.
	The stochastic increments $d\hat B_m(t)=\sqrt{\gamma_m} 
	e^{-\ii\phi_m}~d\hat{ b}_m(t)$ are proportional to the Ito quantum noise
	increments $d\hat{ b}_m$. In the mean-square limit, one can postulate  
	infinitesimal relations for Wiener noise increments
	\begin{align}
	d\hat{ b}_m(t)^{ } d\hat{b}_m^\dagger(t)&=\left(n_m+1\right)dt ,\\
	d\hat{b}_m^\dagger(t) d\hat{b}_m^{ }(t)&=n_mdt ,\\
	d\hat{b}_m(t)d\hat{b}_m(t)&=0.
	\end{align}
	The heat content of the $m$ unsqueezed reservoirs is proportional
	to the occupation numbers $n_m$. 
	Time evolution according to Eq.~\eqref{qsde} is consistent with 
	the Ito rule of calculus   
	\begin{align}
	d(\hat X\hat Y)&\equiv\hat X d\hat Y+(d\hat X)\,\hat Y+d\hat X \,d\hat Y.
	\end{align}
	%Expressions like $d\hat {b}_m(t)dt$ are 
	%of order $\mathcal{O}(dt^{\frac{3}{2}})$ and have to be discarded. 
	%%%%%%%%%%%%%%%%%%%%%%%%%%%%%%%%%%%%%%%%%%%%%%%%%%%%%%%%%
	\section{Multi-channel transmission spectrum of a passive waveguide}
	\label{emptyCavity}
\begin{figure}[h]
		\centering
		\includegraphics[width=1\columnwidth]{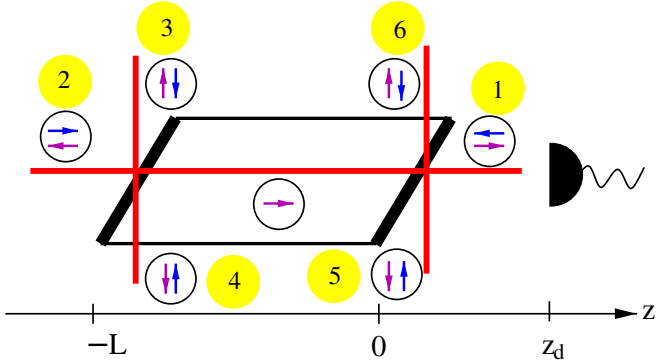}
		\caption{Sketch of the single channels, labeled by numbers between 1 and 6,
			striking the output facets of the QDSLD from the left and right hand side.}
		\label{beamsplitter}
	\end{figure}

	%The geometry of the waveguide or rather the output facets from a QDSLD is of
	%essential importance as it mainly contributes to the characteristic broadband
	%shape of the emitted spectrum. The waveguide is tilted under a certain angle
	%and the facets are anti-reflection coated in order to suppress the formation of
	%longitudinal modes resulting from back reflections at the output ports. This
	%technical feature is modeled by two tilted beam splitters, enclosing the active
	%medium. The single photon detector is located at the right hand side of the
	%diode, as depicted already in Fig.~\ref{detection}. 
	We analyze  the output field measured by the single photon detector at position
	$\vec r_d=(x_d,y_d,z_d)$ under consideration of the facet geometry. For
	simplicity, we study the case of an empty diode, that is in absent of any
	quantum dot or similar gain medium.
	Fig.~\ref{beamsplitter} illustrates the empty diode with the two beam splitters
	modeling the facets, where electrical fields $\hat{\vec E}^\alpha(\vec r,t)$,
	enumerated by the channel numbers $\alpha =1,... ,6$, partially entering and
	simultaneously leaving the diode. 
	% \begin{figure}[h]
	% 	\centering
	% 	\scalebox{.3}{\input{pictures/empty_cavity/beamsplitter.pdf_t}}
	% 	\caption{Sketch of the single channels, labeled by numbers between 1 and 6,
	%striking the output facets of the QDSLD from the left and right hand side.}
	% 	\label{beamsplitter}
	% \end{figure}
	The positive frequency part of the electric field of channel 1 with $z>0$ was
	already introduced and is given by Eq.~\eqref{channelfield}.
	%%%%%%%%%%%%%%%%%%%%%%%%%%%%%%%%%%
	\paragraph{Input-output formalism}
	We apply the input-output formalism \cite{quantumnoise} in order to specify the
	power spectrum of the diode system. The total Hamiltonian depends on the field
	Hamiltonian of the ASE field
	\eqref{fieldhamiltonianase}
	%\begin{align}
	%\hat H_\text{ASE}=\sum_{\{k_i\}} \hbar \omega_i \hat a_i^{\dagger} \hat a_i^{
	%},
	%\end{align}
	the field Hamiltonian of the reservoir corresponding to each channel
	\eqref{hambath}
	%\begin{align}
	%\hat H_b=\sum_{\alpha=1}^6 \hat H_b^{\alpha},\quad	
	%	\hat H_b^{\alpha}=\sum_{\{k_i\}}\sum_{q\in M_i^\alpha} 
	%	\hbar \omega_q \hat b_{i q}^{\alpha\dagger} \hat b_{i q}^{\alpha},
	%\end{align}
	and the Hamiltonian describing the interaction of the external fields with the
	ASE radiation field,
	\begin{align}
	\hat H_{I}=\sum_{\alpha=1}^6\sum_{\{k_i\}}\sum_{q\in M_i^\alpha}
	\ii\hbar \kappa_{ i q}^{\alpha}\hat b_{i q}^{\alpha\dagger}\hat a_i
	+\text{h.c.}.
	\end{align}
	The coupling constant $\kappa_{i q}^{\alpha}=|\kappa_{i q}^{\alpha}|
	\text{exp}(\ii \phi_{i}^{\alpha})$ characterizes the coupling strength between
	the multimode ASE field inside the gain medium and the electrical field
	channels.
	The Heisenberg equations of motion of the system modes reads
	\begin{align}
	\label{bdgl}
	\dot {\hat b}_{i q}^{\alpha}&=-\ii \omega_q
	\hat b_{i q}^{\alpha}
	+\kappa_{i q}^{\alpha} \hat a_i,\\
	\label{adgl}
	\dot {\hat a}_i&= -\ii \omega_i \hat a_i
	- \sum_{\alpha=1}^{6}\sum_{q\in M_i^\alpha}
	\kappa_{i q}^{\alpha\ast}
	\hat b_{i q}^{\alpha}.
	\end{align}
	According to Fig.~\ref{beamsplitter}, channel 3 and 5 
	do not enter the waveguide and can be disregarded. 
	By integrating Eq.~\eqref{bdgl} subject to the initial conditions 
	$\hat{b}_{i q 0}^{\alpha}=\hat{b}_{i q}^{\alpha}(t_0)$, one obtains
	\begin{gather}
	\label{intbath1}
	\hat b_{i q}^{\alpha}(t)=
	e^{-\ii \omega_q(t-t_0)} \hat{b}_{i q0}^{\alpha}
	+\kappa_{i q}^{\alpha} \int\limits_{t_0}^t d\tau e^{-\ii\omega_q(t-\tau)} 
	\hat a_i(\tau),\\
	\dot {\hat a}_i=
	-\ii\omega_i \hat a_i
	-\sum_{\alpha=1}^{6}
	\int\limits_{t_0}^t d\tau
	K_i^{\alpha}(t-\tau)
	\hat a_i(\tau)+ 
	\hat f_i^{\alpha}(t).
	\end{gather}
	This procedure defines stochastic Langevin forces 
	$\hat f_i^\alpha $ and memory kernels $K_i^\alpha$ by
	\begin{align}
	\hat f_i^{\alpha}(t)&
	\equiv\sum_{q\in M_i^\alpha} 
	e^{-\ii\omega_q(t-t_0)}
	\kappa_{i q}^{\alpha\ast} \hat{b}_{i q 0}^{\alpha},\\
	\label{kernel}
	K_i^{\alpha}(\tau) &\equiv 
	\sum_{q\in M_i^\alpha}
	e^{-\ii\omega_q\tau}
	|\kappa_{i q}^{\alpha}|^2.
	\end{align}
	Only for a short time, the stochastic forces are correlated and
	satisfy a two-time fluctuation-dissipation relation
	\begin{align}
	\comut{\hat f_i^{\alpha}(t)}{\hat f_j^{\beta\dagger}(t')}=
	K_i^{\alpha}(t-t')\delta_{\alpha \beta}\delta_{ij}.
	\end{align}
	In the Markov limit, one can approximate convolutions of memory kernels with
	system operators $\hat c(t)$ as
	\begin{align}
	\label{markov}
	\int_{t_0}^t d\tau K_i^{\alpha}(t-\tau)\hat c(\tau)&\approx
	(\tfrac{\gamma_i^{\alpha}}{2}+\ii \delta\omega_i^{\alpha})\hat c(t),
	\end{align}
	for times much larger than the correlation time $\tau_c$, i.e. $t\gg\tau_c$.
	Now, $\gamma_i^\alpha$ defines a decay rate 
	and $\delta\omega^\alpha_i$ is a line shift. Such self-energies implicitly
	renormalize  the bare energies and are not considered furthermore.
	
	The two-time commutation relation of the input field
	\begin{align}
	% \begin{split}
	\comut{\hat b_{i,\text{in}}^{\alpha}(t)}{
		\hat b_{j,\text{in}}^{\beta \dagger}(t')}&=
	\,\delta_{\alpha \beta}\delta_{ij}\delta(t-t'),
	\\
	\delta(\tau)&=\sum_{q\in M_i^\alpha}\tfrac{\Delta\omega_i}{2\pi}
	e^{-\ii\omega_q\tau},
	\nonumber
	% \end{split}
	\end{align}
	has a very short correlation time analog to the Markov approximation 
	\eqref{markov}. In the same order of approximation, the Langevin force,  
	\begin{align}
	\hat f_i^{\alpha}(t)=  \zeta_i^{\alpha\ast} \hat b_{i, \text{in}}^{\alpha}(t)
	\quad
	%\text{with}
	%\quad
	%\zeta_i^\alpha=\sqrt{\gamma_{i}^{\alpha}}
	%e^{\ii \phi_{i}^{\alpha}}
	\end{align} 
	with $\zeta_i^\alpha$ defined in Eq.~\eqref{zeta},
	is proportional to the input operator. 
	Putting all together, we find the equation of motion of the ASE field mode $i$,
	\begin{align}
	%\begin{split}
	\label{ak1}
	\dot{\hat a}_i=
	-\ii \vartheta_i \hat a_i
	-\sum_{\alpha=1}^{6} \hat f_i^{\alpha}(t) ,\quad
	\vartheta_i=\omega_i-\ii\sum_{\alpha=1}^{6}\frac{\gamma_{i}^{\alpha}}{2}
	\end{align}
	in terms of the input operators and damping rates 
	%$\gamma_i^r/2\equiv \gamma_i^1=\gamma_i^2$ and $\gamma_i^l/2\equiv
	%\gamma_i^4=\gamma_i^6$
	\eqref{dampingrate}.
	The corresponding phases are given by 
	%$\phi_i^r\equiv \phi_i^1=\phi_i^2=\phi_i^3$ and $\phi_i^l\equiv
	%\phi_i^4=\phi_i^5=\phi_i^6$
	Eq.~\eqref{phase}, respectively.
	
	Analog calculations in terms of final condition $t_f>t$ lead to a similar
	expression 
	for the ASE field mode,
	\begin{equation}
	\label{ak2}
	\begin{split}
	\dot{\hat a}_i&=-\ii \vartheta_i^\ast \hat a_i
	-\sum_{\alpha=1}^6 \zeta_i^{\alpha\ast} \hat b_{i,\text{out}}^{\alpha}.
	\end{split}
	\end{equation}
	Comparison between Eq.~\eqref{ak1} and Eq.~\eqref{ak2} ends up in the
	fundamental relation,
	\begin{align}
	\label{inputoutputrelation}
	\hat{b}^{\alpha}_{i,\text{out}}(t)
	=\hat b^{\alpha}_{i,\text{in}}(t)+
	\zeta_i^{\alpha}\hat a_i (t),
	\end{align}
	between the input- and output operators. 
	%%%%%%%%%%%%%%%%%%%%%%%%%%%%%%%%%
	\paragraph{Output spectrum}
	The output spectrum recorded by a single-photon sensitive detector with
	cross-section area $A$ corresponds to the expectation value,
	\begin{align}
	S_\text{out}(\omega)=\mathcal C \ave{\hat E^{(-)}(\omega) \hat
		E^{(+)}(\omega)},\quad \mathcal C=2\epsilon _0 c,
	\end{align}
	of the system's output operators in frequency space.  The components of the 
	power spectrum matrix %$S(\omega)$,
	\begin{align}
	\label{spectrumb}
	S_\text{out}(\omega)
	%=\mathcal C\int\limits_A dx dy\ave{\hat{ E}_\text{out}^{\alpha(-)}(\vec
	%r,\omega) 
	%	\hat{ E}_\text{out}^{\beta(+)}	(\vec r,\omega)}
	%\\&
	&=\sum_{\{k_i,k_j>0\}} \Upsilon_{ij} \ave{\hat
		b_{i,\text{out}}^{\alpha\dagger}(\omega) \hat b_{j,\text{out}}^\beta(\omega)},
	\\
	\Upsilon_{ij}&=\frac{2\hbar \pi
		c}{L}\sqrt{\frac{\omega_i\omega_j}{\Delta\omega_i\Delta\omega_j}},
	\end{align}
	depend on the components of $\hat b_{i,\text{out}}(\omega)$ which again can be
	calculated by Fourier transforming the input-output relation
	\eqref{inputoutputrelation} under consideration of the Langevin equation
	Eq.~\eqref{ak2},
	\begin{align}
	\label{vecbout}
	\hat{\vec b}_{i,\text{out}} (\omega)=M_i(\omega)~
	\hat{\vec{b}}_{i,\text{in}}(\omega),
	\end{align}
	with input vector $\hat{\vec b}_{i,\text{in/out}}=(\hat b_{i,\text{in/out}}^{1}
	\dots 
	\hat b_{i,\text{in/out}}^{6} )$. 
	The matrix,
	\begin{align}
	M_{i}& =\frac{1}{\lambda_i}
	\begin{pmatrix}
	m_{i,11}	&  m_{i,12}	& 0	 & m_{i,14} &	0 & m_{i,16}      \\
	m_{i,21}	&  m_{i,22}	& 0	 & m_{i,24} &	0 & m_{i,26}      \\
	0		&	  0		&	 m_{i,33}	 &	 0 	&		0	 	& 	0	      \\
	m_{i,41}	&  m_{i,42}	& 0	 & m_{i,44} &	0 & m_{i,46}      \\
	0				&  	0		& 0	 &		0		 &	m_{i,55}		& 			0      \\
	m_{i,61}		&  m_{i,62}	& 0	 & m_{i,64} &	0 & m_{i,66}      \\
	\end{pmatrix}
	\end{align}
	with $\lambda_i= \ii (\omega-\omega_i)-\gamma_i^{lr}$ and elements
	\begin{align*}
	m_{i,\alpha \beta}&=
	\begin{cases}
	\ii (\omega-\omega_i)+\frac{1}{2}\bigl(\gamma_i^\alpha-\sum_{\alpha'\neq\alpha}
	\gamma_i^{\alpha'}\bigr) & \text{for } \alpha =\beta \\
	\zeta_i^\alpha \zeta_i^{\beta\ast}& \text{for }  \alpha \neq \beta
	\end{cases}
	\end{align*}
	satisfies the relation $M_{i}^\dagger M_{i}^{\phantom\dagger}=1$.
	As depicted in Fig.~\ref{beamsplitter}, only the electrical output field of
	channel $1$ contributes to the power spectrum of the diode, i.e.
	$
	S(\omega)
	=(S_\text{out}(\omega))_{11}
	%=\frac{\hbar c}{\pi L}\sum_{i=1}^N \frac{\omega_i}{\Delta\omega_i} 
	%S_{i,\text{out}}(\omega)_{11}
	.
	$
	Clearly, $S(\omega)$ depends significantly on the choice of the input field: it
	is the convolution of the input field with the response function of the beam
	splitters.
	\paragraph{White noise input}
	Consider the case of vacuum input channel fields $1-5$ and a quantum white noise
	channel $6$ with equal damping rates, $\gamma_i^\alpha=\gamma$. The input matrix
	components reduce to
	%\begin{align}
	%  S_{i,\text{in}}(\omega)_{\alpha \beta}=\begin{cases}
	% 	n &\quad \alpha=\beta=6\\
	% 	0 &\quad  \text{else}
	% \end{cases}
	%\end{align}
	$(S_{ij,\text{in}})_{\alpha\beta}=n^\alpha \delta_{\alpha\beta}\delta_{ij}$
	with $n^6=n=\text{const.}$ and $n^\alpha=0$ else, and the power spectrum
	measured by the detector,
	\begin{align}
	S(\omega)= \sum_{\{k_i>0\}}P_i  \mathcal{L}_\Gamma (\omega-\omega_i) n,
	%	\approx P\sum_{\{k_i>0\}} \mathcal L_i ,
	%	~ 
	%	P_i=\frac{\hbar \omega_i c \gamma_i n}{8\Delta\omega_i L},
	\quad P_i=\frac{ \hbar\pi^2 c \gamma \omega_i}{2\Delta\omega_i L}
	\end{align}
	with $\Gamma=4\gamma$ is given by  the sums of Lorentzians with maximum value at
	$\omega=\omega_i$ and full width at half maximum  $\sqrt{2}\gamma$. 
	Assuming that $\omega_i$ in the enumerator of power $P_i$ differs only slightly
	from a mean value $\bar \omega$, the  continuous power spectrum is approximately
	specified by
	\begin{align}
	S(\omega)=\frac{P}{\Delta\omega} \int_{-\infty}^\infty d\tilde \omega \mathcal
	L_\Gamma(\omega-\tilde\omega)  n.
	%\quad \mathcal E=\frac{2\pi^2 \hbar c \gamma\omega }{L \Delta\omega}.
	\end{align}
	%%%%%%%%%%%%%%%%%%%%%%%%%%%%%%%%%%%%%%%%%%%%%%%%%%
	\section{Fitting the optical power spectrum}
	\label{Fitparam}
	The smooth Gaussian interpolation of the optical power spectrum $S(\omega)$,
	which is depicted in Fig.~\ref{spectrum}, reads
	\begin{align}
	\label{gaussfit}
	S_\text{fit}(\omega)=\frac{S_0 }{\sqrt{2\pi}\sigma}e^{-\frac{(\omega-\bar
			\omega)^2}{2\sigma^2}},
	\quad\int_{-\infty }^\infty d\omega \,S_\text{fit}(\omega)=S_0,
	\end{align}
	with amplitude $S_0$, standard deviation $\sigma$ and central frequency $\bar
	\omega$.
	More generally, one can also propose a Gaussian profile for the cooperativity
	strength,
	\begin{align}
	\label{cooperativity}
	G(\omega )&=G_0 e^{-\frac{(\omega-\bar \omega)^2}{2\sigma^2}},
	\end{align}
	occurring in the quantum model for the intrawaveguide photon number of the QDSLD
	presented in Sec.~\ref{model}. Other parameters, included in this multimode
	theory are assumed to be well known and given by $R=0.5\gamma$,
	$\gamma_{10}=\gamma$, $\gamma_{21}=0.1\gamma$.
	A comparison of  the Gaussian fit of the power spectrum together with the
	solution of the intrawaveguide photon number with Gaussian distributed
	cooperativity strength and equal QDs is depicted Fig.~\ref{datamodel}. The
	obtained fit parameters are  listed in Tab.~\ref{tablefit}. 
	\begin{table}[h!]
		\begin{tabular}{c|| c |c |c|c}
			% \hline
			$\text{Quantity}$ &  $\bar\omega\,\,\, (\si[]{10^{15}/\second})$ & 
			$\sigma\,\,\, (\si[]{10^{12}/\second})$  & $G_0$  (\si{\text{a.u.}}) & $S_0$
			(\si{\text{a.u.}})\\ \hline \hline
			%\hline
			Fit $S_\text{fit}(\omega)$& \num{1.506} & \num{7.286d0} & -- & 0.117 \\
			\hline
			Model $G(\omega)$& \num{1.506} & \num{7.962} & \num{2.467d-4} & --
			% \hline 
		\end{tabular}
		\caption{Fit parameters for the Gaussian and the QDSLD model fit
			corresponding to 
			the power spectrum depicted in Figs.~\ref{spectrum} and ~\ref{datamodel}. The
			normalized sum of the squared residuals are $\sum_{\{k_i\}}
			r_i^2/N=\num{1.137d-4}$ (Gauss) and $\sum_{\{k_i\}} r_i^2/N=\num{0.526d-4}$
			(QDSLD model).} 
		\label{tablefit}
	\end{table}
	%%%%%%%%%%%%%%%%%%%%%%%%%%%%%%%%%%%%%%%%%%%%%%%%%%
	%%%%%%%%%%%%%%%%%%%%%%%%%%%%%%%%%%%%%%%%%%%%%%%%%
%	\bibliographystyle{apsrev4-1}
%	\bibliography{bibliography} 
	\bibliographystyle{apsrev4-1}
	%\bibliography{bibliography} 
	\bibliography{paper2.bbl} 
\end{document}